\documentclass[12pt]{iopart}
\usepackage{iopams}
\usepackage{setstack}

\usepackage{cite}

\usepackage{graphicx}% Include figure files
\usepackage{bm}% bold math
\usepackage{color}

\usepackage[english]{babel}

\newcommand{\rev}[1]{{#1}}
\newcommand{\revneu}[1]{{\it#1}}

\newcommand{\onlinecite}[1]{\cite{#1}}
\newcommand{\vect}[1]{{\bi{#1}}} %vector
\newcommand{\vgr}[1]{{\bm{#1}}} % vectors, tensors of greek letters
\newcommand{\uvec}[1]{{\mathbf{\hat{#1}}}} %unit vectors

\newcommand{\ddfrac}[2]{\frac{\displaystyle #1}{\displaystyle #2}}

\newcommand{\rh}[1]{\rho^{(#1)}}

\newcommand{\curr}[1]{\vgr{\mathcal{J}}_{\!#1}} % current density

\usepackage{xcolor}
\begin{document}

\noindent{\revneu{Original content from this
work may be used under
the terms of the Creative
Commons Attribution 3.0
licence (https://creativecommons.org/licenses/by/3.0/legalcode).
Any further distribution of
this work must maintain
attribution to the
author(s) and the title of
the work, journal citation
and DOI.}\\
\revneu{To cite this article: {\normalfont Christian Hoell} {\it et al} {\normalfont 2017} {\it New J. Phys.} {\bf 19}, {\normalfont 125004}.\\
This is an author-created, un-copyedited version of an article published 
in New Journal of Physics. IOP Publishing Ltd is not responsible for any errors or omissions in this version of the manuscript or 
any version derived from it. The Version of Record is available online at 
https://doi.org/10.1088/1367-2630/aa942e.}
\\

\title{Dynamical density functional theory for circle swimmers}

\author{Christian Hoell}

\author{Hartmut L\"owen}

\author{Andreas M.~Menzel*}
% \mailto{*mail:menzel@hhu.de}
\address{Institut f\"ur Theoretische Physik II, Weiche Materie,
Heinrich-Heine-Universit\"at D\"usseldorf,
40225 D\"usseldorf, Germany.}

\date{\today}

\begin{abstract}
The majority of studies on self-propelled particles and microswimmers concentrates on objects that do not feature a deterministic bending of their trajectory. However, perfect axial symmetry is hardly found in reality, and shape-asymmetric active microswimmers tend to show a persistent curvature of their trajectories. Consequently, we here present a particle-scale statistical approach of circle-swimmer suspensions in terms of a dynamical density functional theory. It is based on a minimal microswimmer model and, particularly, includes hydrodynamic interactions between the swimmers. After deriving the theory, we numerically investigate a planar example situation of confining the swimmers in a circularly symmetric potential trap. There, we find that increasing curvature of the swimming trajectories can reverse the qualitative effect of active drive. More precisely, with increasing curvature, the swimmers less effectively push outwards against the confinement, but instead form high-density patches in the center of the trap. We conclude that the circular motion of the individual swimmers has a localizing effect, also in the presence of hydrodynamic interactions. Parts of our results could be confirmed experimentally, for instance, using suspensions of L-shaped circle swimmers of different aspect ratio. 
\end{abstract}


\maketitle

\section{Introduction}

On the scales of active colloidal particles and self-propelled biological microswimmers \cite{berg1972chemotaxis,paxton2004catalytic,howse2007self, lauga2009hydrodynamics,romanczuk2012active,cates2012diffusive, menzel2015tuned,elgeti2015physics,zottl2016emergent, bechinger2016active}, thermal fluctuations and other perturbations play a prominent role. They lead to continuous reorientation of the self-propelling objects and therefore to stochastically shaped trajectories \cite{paxton2004catalytic,howse2007self,buttinoni2012active}. Even more extreme events are given by stochastic run-and-tumble motions. For instance, certain bacteria
or \rev{alga} cells
are observed to stop their migration, reorient basically on the spot, and then continue their propulsion \cite{berg1972chemotaxis,polin2009chlamydomonas}. Such events lead to kinks on the trajectory. The statistics of both types of buckled motion has been studied in detail, both in experiment and in theory \cite{berg1972chemotaxis,howse2007self,tailleur2008statistical, min2009high, ten2011brownian,volpe2011microswimmers,buttinoni2012active, bennett2013emergent,zheng2013non,sevilla2016diffusion}. Yet, in the absence of any noise, fluctuations, and perturbations, the self-propelling agents considered in most theoretical analyses would show a deterministic straight motion. 

Here, we concentrate on active microswimmers that feature a different behavior. Their individual trajectories are systematically curved. Such a situation can arise only, if for each swimmer the axial symmetry around its propulsion direction is broken. 

On the one hand, the symmetry breaking can be induced from outside. For instance, if microswimmers are exposed to local surrounding shear flows, the rotational component of the fluid flow can couple to the orientation of the suspended swimmer \cite{ten2011brownian,zottl2012nonlinear,tarama2013dynamics, tournus2015flexibility,rusconi2015microbes, mathijssen2016upstream,mathijssen2016hotspots}. Continuous reorientation of the propulsion direction leads to curved trajectories. Similarly, the symmetry is broken in the presence of a nearby surface. If during propulsion a swimmer shows rotations of its body around its axis, these rotations can on one side hydrodynamically interact with the surface. Via such hydrodynamic surface interactions the self-rotation couples to the propulsion direction and the trajectory bends. Also steric interactions can support or induce the effect. Thus circular trajectories are observed for many sperm cells and bacteria close to a substrate \cite{ramia1993role,frymier1995three,lauga2006swimming, riedel2005self,elgeti2010hydrodynamics}. 

On the other hand, real swimmers often bring along a broken axial symmetry by themselves \cite{wittkowski2012self}. Hardly any object is really perfectly axially symmetric in shape. On purpose, L-shaped active microswimmers have been fabricated and their persistently curved trajectories were analyzed \cite{ledesma2012circle,kummel2013circular,hagen2014gravitaxis, ten2015can}. If the trajectories, including their persistent bending, are confined to a plane, then circular paths arise. This is what we understand by circle swimmers \cite{lowen2016chirality}. Apart from that, for deformable self-propelled particles and self-propelled nematic droplets, the symmetry breaking in shape or structure may also occur spontaneously \cite{ohta2009deformable,hiraiwa2011dynamics,kruger2016curling}. Moreover, imperfections in the self-propulsion mechanism can lead to the symmetry breaking and thus to bent trajectories. An example are cells of the algal \textit{Chlamydomonas reinhardtii}. If one of the two beating flagella generating self-propulsion is weaker or absent, the cellular paths curve \cite{brokaw1982analysis,kamiya1984submicromolar}. Apart from that, near surfaces bent self-propelled objects tend to follow circular trajectories \cite{takagi2013dispersion,denk2016active}. In modeling approaches, circle swimmers are often realized by simply imposing an effective torque or rotational drive in addition to the self-propulsion mechanism \cite{van2008dynamics,van2009clockwise,ten2011brownian, weber2011active,wittkowski2012self,fily2012cooperative, radtke2012directed,weber2012active,kaiser2013vortex, mijalkov2013sorting,reichhardt2013dynamics,marine2013diffusive, yang2014self,chen2015sorting, nourhani2015guiding,ten2015can,ao2015diffusion, liebchen2016rotating,jahanshahi2017brownian}. 

We have mentioned above that studies on circle swimmers are relatively rarely encountered when compared to the numbers of works on objects propelling straight ahead. Even less frequent are studies on the collective behavior of circle swimmers \cite{riedel2005self,kaiser2013vortex,reichhardt2013dynamics, yang2014self,denk2016active}. Particularly, this applies when hydrodynamic interactions in crowds of suspended microswimmers are to be included. 

When the collective properties of many interacting agents are investigated, such statistical approaches become important  \cite{baskaran2009statistical,menzel2012collective,chou2012kinetic,grossmann2015pattern,chou2015active,heidenreich2016hydrodynamic,menzel2016dynamical}. Recently, we have derived and evaluated a microscopic statistical description for straight-propelling microswimmers in terms of a classical dynamical density functional theory (DDFT) \cite{menzel2016dynamical}. Microscopic here means that the description is based and operates on the length scales of the individual agents. Thus, for instance, when classical density functional theory or its variants are used to describe the properties of crystalline structures \cite{elder2002modeling,elder2004modeling,goldenfeld2005renormalization, elder2007phase,van2008colloidal, van2009derivation,tegze2011faceting,neuhaus2013compatibility, menzel2013traveling,van2013vacancy, menzel2014active,archer2014solidification, chervanyov2016effect,zimmermann2016flow}, individual crystal peaks can be resolved in the statistical density field. 

In equilibrium, i.e., for passive systems, density functional theory \cite{hansen1990theory,singh1991density,evans1992density, evans2010density,lowen2010density} is, in principle, an exact theory. It can be extended to overdamped relaxational dynamics in terms of DDFT \cite{marconi1999dynamic,marconi2000dynamic,archer2004dynamical, lowen2010density} by assuming at each instant an effective equilibrium situation to evaluate the involved potential interactions. For example, solidification processes are addressed in this way \cite{van2008colloidal,van2009derivation, neuhaus2013compatibility,archer2014solidification}. Since microswimmers by construction operate at low Reynolds numbers \cite{purcell1977life}, their dynamics is overdamped. This makes DDFT a promising candidate to study their statistical behavior. 

Extending density functional theory to intrinsically non-equilibrium systems, DDFTs for ``dry'' self-propelling agents had already been developed before \cite{wensink2008aggregation,wittkowski2011dynamical,pototsky2012active} and tested against agent-based simulations \cite{wensink2008aggregation,pototsky2012active}. Moreover, to characterize passive colloidal particles in suspensions, hydrodynamic interactions had been incorporated into DDFT \cite{rex2008dynamical,rex2009dynamical,rauscher2010ddft, goddard2012general,goddard2012unification,donev2014dynamic,goddard2016dynamical} and agreement was found with explicit particle-based simulations \cite{rex2008dynamical,rex2009dynamical, goddard2012general,goddard2012unification}. Our recently developed DDFT for microswimmers incorporates and combines all the central previous ingredients, i.e., self-propulsion, steric interactions between the swimmers, hydrodynamic interactions between the swimmers, as well as exposure to and confinement by external potentials \cite{menzel2016dynamical}. As we have demonstrated and as further detailed below, this dynamical theory qualitatively reproduces previous simulation results \cite{nash2010run,hennes2014self} in which combined action of all these ingredients determines the overall behavior. 

Here, we proceed by an additional step forward. We extend our microscopic statistical characterization (DDFT) to circle swimmers. In this way, we can now characterize the collective behavior of such non-straight-propelling agents, including the effect of hydrodynamic interactions. Only then, for instance and as we will show below, can the symmetry breaking induced by hydrodynamic interactions in a radial confinement be described qualitatively correctly.

We first introduce our minimal model for circle swimmers in section \ref{sec:Model}. Next, in section \ref{sec:DDFT}, we list the extension of the theory. It is evaluated numerically in section \ref{sec:trap} to study the behavior of circle swimmers under radial confinement as a function of the bending of their trajectory. 
A short summary and outlook are given in section \ref{sec:Conclusions}.

\section{Minimal model circle swimmer}
\label{sec:Model}

As outlined above, our goal is to establish a microscopic DDFT of circle-swimmer suspensions. The term `microscopic' here refers to the length scales of an individual microswimmer. To base our DDFT on such scales, we need to first introduce an explicit minimal model for a microscopic circle swimmer. 

\begin{figure}
\centering
\includegraphics[width=0.75\linewidth,trim=0 0 0 5pt, clip=true]{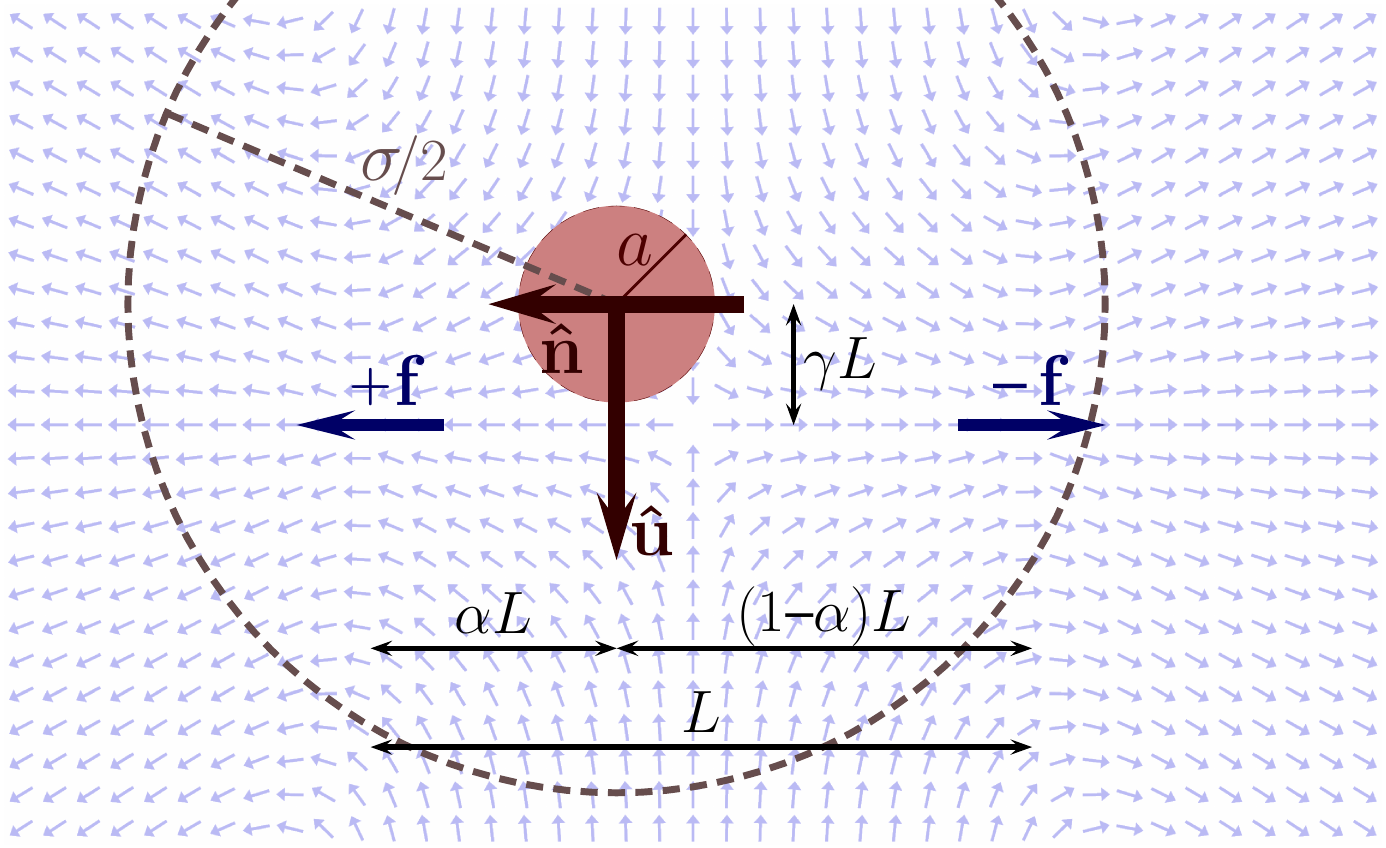}
\caption{Minimal model of a circle microswimmer. Two active force centers separated by a distance $L$ exert the active forces $\pm\vect{f} = \pm f \uvec{n}$ onto the fluid, where $\uvec{n}$ marks the direction of the principal swimmer axis. The resulting fluid flow is indicated by the small light arrows that are rescaled to identical length for visualization. A spherical swimmer body of effective hydrodynamic radius $a$ is, in general, asymmetrically placed between the force centers. Its shift along $\uvec{n}$ out of the symmetry center, quantified by $\alpha\neq1/2$, leads to forward (or backward) propulsion. An additional transversal shift, quantified by $\gamma$, implies circular trajectories for $\alpha\neq1/2$ and $\gamma\neq0$. This circle swimmer is \rev{biaxial}, the secondary axis denoted by $\uvec{u}$. We impose spherically symmetric steric interactions between the swimmers, with an effective interaction radius $\sigma/2$.}
\label{fig:model}
\end{figure}
\rev{Our statistical theory will apply to (semi)dilute suspensions of microswimmers based on their far-field hydrodynamic interactions. Therefore, a minimal model microswimmer is needed that shows the correct leading-order far-field hydrodynamic fluid flows together with a self-consistent description of its self-propulsion. Yet, at the same time, it must be simple enough to still be efficiently included into the statistical description.} Figure~1 represents our corresponding swimmer model. The unit vector $\uvec{n}$ identifies the principal swimmer axis and orientation. 

\rev{Any active microswimmer exerts forces onto the surrounding fluid. Amongst them are the spatially distributed active forces that initiate self-propulsion. They are generated, for instance, by the rotation of flagella or beating of cilia \cite{polin2009chlamydomonas, min2009high, brokaw1982analysis, kamiya1984submicromolar}. In our model, all these active forces are thought to be gathered and concentrated in one spot. In figure~1, this leads to the active point force $-\vect{f}$ acting on the surrounding fluid. Instantly, due to the nature of the considered low-Reynolds-number motion, see below \cite{purcell1977life}, for a freely suspended microswimmer all these active forces are balanced by frictional forces distributed over its body. We consider all these counteracting frictional forces to be concentrated in another spot, leading to the point force $+\vect{f}$ in figure~1. These two spots in general do not coincide, depending on the actual swimmer geometry. Here, they are separated by a distance $L$, see figure~1. However, since no net force nor torque may act on a freely suspended microswimmer, the two forces $\pm\vect{f}$ need to be of same magnitude but oppositely oriented, located and aligned along a common axis. We may thus parameterize them as $\pm\vect{f}=\pm f\uvec{n}$. They act} onto the fluid and set it into motion as indicated by the small arrows in the background \rev{of figure~1}. In analogy to straight-swimming terminology \rev{\cite{lauga2009hydrodynamics}}, for $f>0$, i.e., the depicted case, we call the object a \textit{pusher}. For $f<0$, we term it a \textit{puller}. 

Next, we place a spherical swimmer body of effective hydrodynamic radius $a$ \rev{nearby} the two force centers. \rev{The whole construct is a rigid object, i.e., the force centers and forces have to rigidly translate and rotate together with the sphere, maintaining their mutual distances and orientations. The role of the sphere is purely to realize self-propulsion of the whole object. Since all forces exerted by the swimmer onto the fluid have already been concentrated into the two force centers above (ignoring all forces that lead to higher-order contributions to the hydrodynamic far-field) the sphere is considered not to exert any remaining force onto the fluid any longer. Its sole role is to be convected by the self-induced fluid flow, leading to the overall self-propulsion.} Unless it is positioned into the \rev{exact point} of symmetry \rev{between the two force centers}, a net transport of the \rev{swimmer} results in the induced fluid flow. For a shift of the \rev{sphere} along $\uvec{n}$ out of the symmetry plane between the two force centers, the whole object propels into the direction of one of the two forces. This shift is quantified by the parameter $\alpha$, with $\alpha=1/2$ marking the symmetric configuration. 

In addition to our swimmer model in Ref.~\onlinecite{menzel2016dynamical}, we now consider an extra shift of the \rev{spherical} swimmer body into a direction perpendicular to $\uvec{n}$. The parameter $\gamma$ quantifies this shift, see figure~1, so that the axial symmetry is broken for $\gamma\neq0$. Consequently, for $\alpha\neq1/2$ and $\gamma\neq0$, the swimmer in the absence of any fluctuations starts to circle, as quantified below. Moreover, it is now \rev{biaxial}, with the additional axis marked as $\uvec{u}$, see figure~1. 

\rev{Since we consider the} hydrodynamic interactions \rev{at a far-field level, we need to} hinder the microswimmers from coming too close to each other. \rev{Therefore,} we consider spherically symmetric soft steric interactions between the swimmer bodies of effective radius $\sigma/2>[(\mathrm{max}\{\alpha,1-\alpha\})^2+\gamma^2]^{1/2}L$ \rev{to maintain an effective distance between them}. 
Altogether, the whole rigid swimmer object in figure~1 is force- and torque-free, as mandatory for a microswimmer suspended in a bulk fluid, see also Appendix A.''

\subsection{Hydrodynamic interactions}

We now consider $N$ identical circle microswimmers suspended in the fluid and use indices $i=1,...,N$ to label them. As described above, for $f\neq0$, each circle swimmer sets the surrounding fluid into motion due to its active forces exerted by the active force centers. In addition to that, the swimmer bodies may be subjects to forces $\vect{F}_i$ and torques $\vect{T}_i$. These may, for instance, be stochastic in nature, result from steric interactions between the circle swimmers, or be imposed from outside. 
Since the dynamics of microswimmers is usually determined by low Reynolds numbers \cite{purcell1977life}, it is described by the linear Stokes equation \cite{Dhont}. That is, their dynamics is overdamped, and the forces $\vect{F}_i$ and torques $\vect{T}_i$ are directly transmitted to the surrounding fluid, setting it into motion. Moreover, since the swimmers are suspended in the fluid, they are translated and rotated by the induced fluid flows. The instantly resulting velocities $\vect{v}_i$ and $\vgr{\omega}_i$ are calculated from a matrix equation as \cite{menzel2016dynamical}
\begin{equation}
\left[
\begin{array}{c}
 \vect{v}_i \\[0.1cm]
 \vgr{\omega}_i
\end{array}
\right]
=
    \sum_{j=1}^{N}
    \left(
\left[
    \begin{array}{cc}
    \vgr{\mu}^{\mathrm{tt}}_{ij} & \vgr{\mu}^{\mathrm{tr}}_{ij}\\[0.1cm]
    \vgr{\mu}^{\mathrm{rt}}_{ij} & \vgr{\mu}^{\mathrm{rr}}_{ij}    
    \end{array}
\right]
    \cdot
\left[
    \begin{array}{c}
        \vect{F}_j \\[0.1cm]
        \vect{T}_j
    \end{array}
\right]
    +
\left[
    \begin{array}{cc}
    \vgr{\Lambda}^{\mathrm{tt}}_{ij} & \vect{0}\\[0.1cm]
    \vgr{\Lambda}^{\mathrm{rt}}_{ij} & \vect{0}   
    \end{array}
\right]
    \cdot
\left[
    \begin{array}{c}
        f \uvec{n}_j \\[0.1cm]
        \vect{0}
    \end{array}
\right]
    \right)
  \label{mobility}
\end{equation}
for $i=1,...,N$. 

In (\ref{mobility}), the first product on the right-hand side includes the influence of the passive swimmer bodies. 
$\vgr{\mu}^{\mathrm{tt}}_{ij}$, $\vgr{\mu}^{\mathrm{tr}}_{ij}$, $\vgr{\mu}^{\mathrm{rt}}_{ij}$, and $\vgr{\mu}^{\mathrm{rr}}_{ij}$ are the familiar mobility matrices that express how swimmer $i$ is translated and rotated due to the forces and torques transmitted by the swimmer body $j$ onto the fluid \cite{Rotne_1969_JCP,Dhont,reichert2004hydrodynamic, menzel2016dynamical}. These expressions are the same as for suspended passive colloidal particles and result from an expansion in the inverse separation distance between the swimmer bodies, where here we proceed up to the third order, i.e., the Rotne-Prager level. 

Then, for $i=j$, we have \cite{Rotne_1969_JCP,Dhont,reichert2004hydrodynamic, menzel2016dynamical}
\begin{eqnarray}
\vgr{\mu} ^{\mathrm{tt}}_{ii}&=&\mu^{\mathrm{t}}{\mathbf{1}},
\quad\vgr{\mu}_{ii}^{\mathrm{rr}}=\mu^{\mathrm{r}}{\mathbf{1}},
\quad\vgr{\mu}_{ii}^{\mathrm{tr}}=\vgr{\mu}_{ii}^{\mathrm{rt}}=\vect{0},
\label{Stokes2}
\end{eqnarray} 
with
\begin{equation}\label{abbr}
\mu^{\mathrm{t}}=\frac{1}{6\pi\eta a}, \qquad \mu^{\mathrm{r}}=\frac{1}{8\pi\eta a^3}.
\end{equation}
Here, $\eta$ is the viscosity of the fluid. 

For $i\neq j$, the mobility matrices read \cite{Rotne_1969_JCP,Dhont,reichert2004hydrodynamic, menzel2016dynamical}
\begin{eqnarray}
\vgr{\mu}_{ij}^{\mathrm{tt}}&=&\mu^{\mathrm{t}}\Bigg(\frac{3a}{4r_{ij}}\Big({\mathbf{1}}+{\uvec{r}_{ij}\uvec{r}_{ij}}\Big) 
+\frac{1}{2}\left(\frac{a}{r_{ij}}\right)^{\!\!3}\Big({\mathbf{1}}-3{{\uvec{r}_{ij}\uvec{r}_{ij}}}\Big)\Bigg),
\label{mu_tt}
\\
\vgr{\mu}_{ij}^{\mathrm{rr}}&=&{}-\frac{1}{2}\,\mu^{\mathrm{r}}\left(\frac{a}{r_{ij}}\right)^{\!\!3}\left({\mathbf{1}}-3{{\uvec{r}_{ij}\uvec{r}_{ij}}}\right), \\ 
\vgr{\mu}_{ij}^{\mathrm{tr}}&=&\vgr{\mu}_{ij}^{\mathrm{rt}}=\mu^{\mathrm{r}}\left(\frac{a}{r_{ij}}\right)^{\!\!3}{ {\vect{r}_{ij}}}\times, 
\label{mu_tr}
\end{eqnarray}
where $\vect{r}_{ij}=\vect{r}_j-\vect{r}_i$, with $\vect{r}_i$ and $\vect{r}_j$ marking the swimmer positions, 
$r_{ij}=|\vect{r}_{ij}|$, 
$\uvec{r}_{ij}=\vect{r}_{ij} / r_{ij}$, 
and ``$\times$'' is the vector product. 

The second product on the right-hand side of (\ref{mobility}) arises because of the active forces that the swimmers exert onto the fluid. Naturally, these actively induced fluid flows likewise contribute to the velocities $\vect{v}_i$ and angular velocities $\vgr{\omega}_i$ of all swimmer bodies. 
$\vgr{\Lambda}^{\mathrm{tt}}_{ij}$ and $\vgr{\Lambda}^{\mathrm{rt}}_{ij}$ are the corresponding mobility matrices. 
The entries $\vect{0}$ in these expressions arise because our swimmers do not exert active torques onto the suspending fluid.

More precisely, the mobility matrices $\vgr{\Lambda}^{\mathrm{tt}}_{ij}$ and $\vgr{\Lambda}^{\mathrm{rt}}_{ij}$ describe how the active forces exerted by the two force centers of swimmer $j$ onto the fluid influence the velocity $\vect{v}_i$ and angular velocity $\vgr{\omega}_i$ of swimmer $i$, respectively. Since swimmer $j$ carries two active force centers exerting the two forces $\pm\vect{f}_j=\pm f\uvec{n}_j$, both $\vgr{\Lambda}^{\mathrm{tt}}_{ij}$ and $\vgr{\Lambda}^{\mathrm{rt}}_{ij}$ split into two contributions \cite{menzel2016dynamical},
\begin{eqnarray}
\vgr{\Lambda}_{ij}^{\mathrm{tt}} &=& \vgr{\mu}_{ij}^{\mathrm{tt}+}-\vgr{\mu}_{ij}^{\mathrm{tt}-},
\label{Lambda_tt} \\
\vgr{\Lambda}_{ij}^{\mathrm{rt}} &=& \vgr{\mu}_{ij}^{\mathrm{rt}+}-\vgr{\mu}_{ij}^{\mathrm{rt}-}. 
\label{Lambda_rt}
\end{eqnarray}
In contrast to the passive swimmer bodies, the active force centers are point-like. 
Therefore, the expressions for the four mobility matrices $\vgr{\mu}_{ij}^{\mathrm{tt}\pm}$ and $\vgr{\mu}_{ij}^{\mathrm{rt}\pm}$ are slightly modified when compared to the corresponding expressions for the hydrodynamic interactions between the passive  
swimmer bodies in (\ref{mu_tt}) and (\ref{mu_tr}) \cite{menzel2016dynamical},
\begin{eqnarray}
\vgr{\mu}_{ij}^{\mathrm{tt}\pm} &=&
\frac{1}{8\pi\eta r_{ij}^{\pm}}\left({\mat {1}}+\uvec{r}_{ij}^{\pm}\uvec{r}_{ij}^{\pm}\right) 
+\frac{a^2}{24\pi\eta {\left(r_{ij}^{\pm}\right)}^{\!3}}\left({\mat {1}}-3\uvec{r}^{\pm}_{ij}\uvec{r}^{\pm}_{ij}\right), 
\nonumber\\[-.2cm]&&
\label{mu_tt_pm}\\
\vgr{\mu}_{ij}^{\mathrm{rt}\pm} &=& \frac{1}{8\pi\eta \left({r_{ij}^{\pm}}\right)^{\!3}} \vect{r}_{ij}^{\pm}\times.
\label{mu_rt_pm}
\end{eqnarray} 
Here, $\vect{r}_{ij}^{\pm}$ are the distance vectors between the passive body of swimmer $i$ and the active force centers of swimmer $j$, exerting the forces $\pm\vect{f}_j=\pm f\uvec{n}_j$ onto the fluid, respectively. Again, $r_{ij}^{\pm}=|\vect{r}_{ij}^{\pm}|$ and $\uvec{r}_{ij}^{\pm}=\vect{r}_{ij}^{\pm}/r_{ij}^{\pm}$. 
In contrast to Ref.~\onlinecite{menzel2016dynamical}, where straight-propelling microswimmers were investigated, we here must take into account the additional transversal shift of the active force centers with respect to the swimmer bodies, see figure~\ref{fig:model}. Therefore, we now obtain
\begin{eqnarray}
\vect{r}_{ij}^+ &=& \vect{r}_{ij}+\alpha L \uvec{n}_j + \gamma L \uvec{u}_j, \label{defplus}
\\
\vect{r}_{ij}^- &=& \vect{r}_{ij}-(1-\alpha) L \uvec{n}_j + \gamma L \uvec{u}_j. \label{defminus}
\end{eqnarray}
Naturally, the values of $\alpha$ and $\gamma$ must assure that the force centers of each swimmer are located outside the hydrodynamic radius $a$ of the swimmer body, i.e., $[(\mathrm{min}\{\alpha,1-\alpha\})^2+\gamma^2]^{1/2}L>a$.  

We consider our spherical swimmer body to exclusively act as a probe particle. Therefore our active mobility matrices for interactions between different swimmers ($i\neq j$) are given to lowest order in $(\alpha-0.5)$ and/or $a/L$. In a next step, the distortion of the flow field by the rigid swimmer body could be included by considering the image system within a rigid sphere \cite{Kim_Karrila, adhyapak2017flow}.

Moreover, for $i=j$, (\ref{mobility}) together with (\ref{Lambda_tt})--(\ref{defminus}) describe the self-propelled motion of one individual circle swimmer. 
At the moment not considering any fluctuations, one such isolated microswimmer ($N=1$) keeps self-propelling with constant translational speed $v_\mathrm{s}$ and constant angular speed $\omega_\mathrm{s}$ along a closed circular trajectory of radius $R_\mathrm{s}=v_\mathrm{s} / \omega_\mathrm{s}$ for all times. 
Since both $v_\mathrm{s}$ and $\omega_\mathrm{s}$ depend on the relative position between the swimmer body and the two force centers, $R_\mathrm{s}$ can smoothly be tuned between almost zero and infinity by altering the parameters $\alpha$ and $\gamma$; see figure~\ref{fig:circle_radius}. Moreover, both $\vgr{\Lambda}^{\mathrm{tt}}_{ii}$ and $\vgr{\Lambda}^{\mathrm{rt}}_{ii}$ 
are independent of $f$. Thus, both $v_\mathrm{s}$ and $\omega_\mathrm{s}$ scale linearly with $f$, see (\ref{Lambda_tt})--(\ref{defminus}). Therefore, $R_\mathrm{s}$ is independent of the active force $f$. Swimming faster does not change the radius of the circle. 

Technically, our mobility matrices represent the solutions to the underlying Stokes equation for the flow of the suspending fluid at low Reynolds number \cite{Dhont}. In this way, the role of the fluid is implicitly included in our description.

\begin{figure}
 \centering
\includegraphics[width=0.75\linewidth]{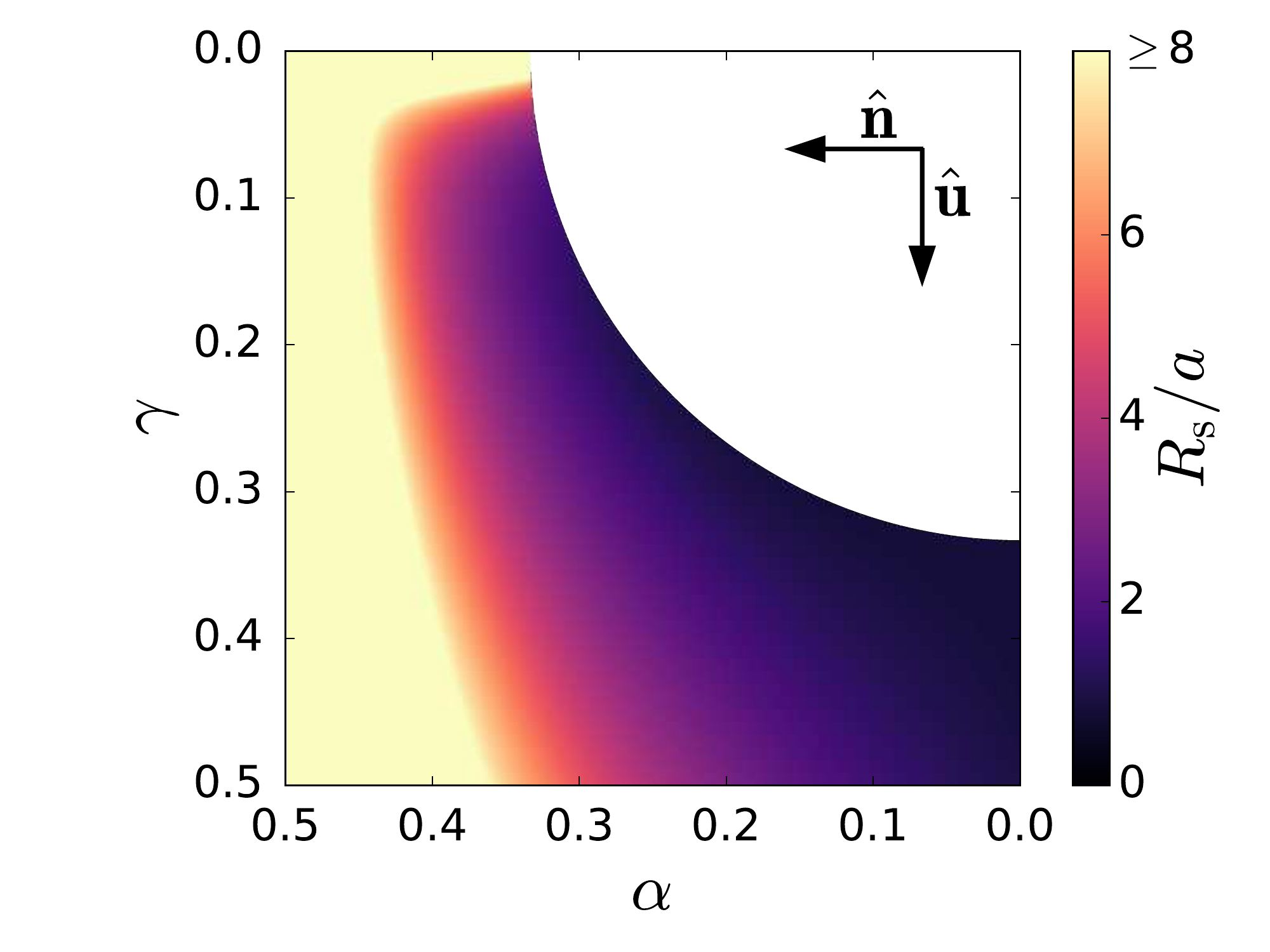}
 \caption{
 Radius $R_\mathrm{s}$ of the circular trajectory at vanishing fluctuations for one single isolated circle microswimmer as introduced in figure~\ref{fig:model} ($L/a=3$). The color map indicates $R_\mathrm{s}$ as a function of $\alpha$ and $\gamma$. For $\gamma \rightarrow 0$, $R_\mathrm{s}/a\rightarrow\infty$ and the swimmer self-propels straight ahead. 
We do not allow a force center to be placed within the hydrodynamic radius $a$ from the center of the swimmer body, reflected by the white area on the top right.}
\label{fig:circle_radius}
\end{figure}

\subsection{Stochastic forces, external forces, and steric interactions}

Our remaining task is to specify the forces $\vect{F}_i$ and torques $\vect{T}_i$ acting on the swimmer bodies in (\ref{mobility}). 
The forces are set to 
\begin{equation}
\vect{F}_i =  {}-k_\mathrm{B} T \; {\nabla_i\ln P}-\nabla_i U.
\label{force}
\end{equation}
Here, the first contribution represents the effective influence of the stochastic forces due to thermal fluctuations \cite{doi1988theory}. $k_B$ is the Boltzmann constant, $T$ the temperature, $\nabla_i=\partial/\partial\vect{r}_i$, and $P=P(\vect{r}_1,\uvec{n}_1,\uvec{u}_1,\dots,\vect{r}_N,\uvec{n}_N,\uvec{u}_N,t)$ is the probability density to find at a certain time $t$ the swimmers at positions $\vect{r}_i$ with orientations $\uvec{n}_i$ and $\uvec{u}_i$, $i=1,\dots,N$. From this form, the correct diffusional behavior is reproduced in the statistical approach, see below. 

The overall potential in the second part of (\ref{force}) reads
\begin{equation}
U=\frac{1}{2}\sum_{\substack{k,l=1;\, k\neq l}}^Nu(\vect{r}_k,\vect{r}_l)+\sum_{k=1}^N u_\mathrm{ext}(\vect{r}_k).
\label{eq_U}
\end{equation}
In this expression, the first term describes the steric interactions between the swimmer bodies. We here choose a soft GEM-4 potential of the form \cite{mladek2006formation,archer2014solidification}
\begin{equation}
u(\vect{r}_k,\vect{r}_l)=\epsilon_0\exp\left(-\frac{r_{kl}^4}{\sigma^4}\right), 
\label{eqGEM4}
\end{equation}
where $\epsilon_0$ sets the strength of the interactions. $u_\mathrm{ext}$ is an external potential acting on each swimmer body and further addressed below. 

Finally, the only torques that we consider to act on our spherically symmetric swimmer bodies are stochastic ones, 
\begin{eqnarray}
{\vect{T}_i} = -k_\mathrm{B} T \; \nabla^\mathrm{or}_i \ln P.
\label{torque}
\end{eqnarray} 
Here, the operator $\nabla^\mathrm{or}_i$ contains the derivatives with respect to the particle orientations. If the swimmers and their orientations are confined to a flat plane, for instance, the $xy$ plane in Cartesian coordinates, one angle $\varphi_i$ is sufficient to characterize the orientation of each swimmer $i$. Then the operator reduces to $\nabla^\mathrm{or}_i=\uvec{z}\,\partial / \partial \varphi_i$\rev{, where $\uvec{z}$ is the (oriented) Cartesian unit vector perpendicular to the $xy$ plane in a three-dimensional Euclidean space.}
In three dimensions, explicit expressions using Eulerian angles exist \cite{Gray_Gubbins, wittkowski2011dynamical}.

\section{Dynamical density functional theory for circle swimmers}
\label{sec:DDFT}

Based on our minimal microswimmer model, we can now derive a microscopic statistical description in terms of a dynamical density functional theory (DDFT) for suspensions of identical circle swimmers. The derivation follows the same lines as in our previous work on straight-propelling microswimmers \cite{menzel2016dynamical}. However, several changes result from the present \rev{biaxiality} of the individual swimmers.

We start from the microscopic Smoluchowski equation
\begin{equation}
\frac{\partial P}{\partial t}= -\sum_{i=1}^N\Big(\nabla_i\cdot\left(\vect{v}_i P\right)
+
\nabla^\mathrm{or}_i
\cdot\left(\vgr{\omega}_i P\right)\Big),
\label{Smoluchowski}
\end{equation}
which states the conservation of the overall probability density. Here, we have to insert the swimmer velocities $\vect{v}_i$ and angular velocities $\vgr{\omega}_i$ as given by (\ref{mobility})--(\ref{torque}). 
\rev{Although $\vect{v}_i$ and $\bm{\omega}_i$ depend on $\ln P$ via (13) and (16), it is important to stress that (17) is still linear in $P$.
Using the chain rule in (13) leads to $\nabla_i \ln P = (\nabla_i P) / P$, which in combination with the factor $P$ in (17) leads to the linear contribution $\nabla_i P$.
The same argument applies to the term $\nabla^\mathrm{or}_i$ in (16) when inserted into (17).}

To obtain from (\ref{Smoluchowski}) the $n$-swimmer density of finding $n$ of the identical $N$ circle swimmers at a certain time at certain positions with certain orientations, we must integrate out from (\ref{Smoluchowski}) all but the degrees of freedom of $n$ swimmers. We denote by $\vect{X}_i$ 
all degrees of freedom of the $i$th swimmer. Then, the $n$-swimmer density is obtained from the overall probability density $P$ as
\begin{equation}
\rho^{(n)}(\vect{X}_1,\dots,\vect{X}_n,t) = \frac{N!}{(N-n)!}
\int \mathrm{d} \vect{X}_{n+1} \dots \mathrm{d} \vect{X}_{N}
\;P.
\label{nbody}
\end{equation}
In the special case of all swimmers and their orientations being confined to a flat plane, $\vect{X}_i=(\vect{r}_i, \varphi_i)$ and $\mathrm{d} \vect{X}_i=\mathrm{d}\vect{r}_i \mathrm{d} \varphi_i$.

Our goal is to obtain an equation for the dynamics of the one-swimmer density $\rho^{(1)}(\vect{X},t)$ to find a circle swimmer at time $t$ with position and orientation(s) $\vect{X}$. However, the integration scheme in (\ref{nbody}) leads to a non-closed equation for the time derivative of $\rho^{(1)}$. Because of our pairwise hydrodynamic and steric interactions, $\rho^{(1)}$ couples to the pair density $\rho^{(2)}$, and, in combination of both interactions, also to $\rho^{(3)}$ \cite{rex2008dynamical,rex2009dynamical,menzel2016dynamical}. This starts a whole hierarchy of coupled dynamical equations, called BBGKY hierarchy \cite{hansen1990theory}. 
To close the dynamical equation for $\rho^{(1)}$, we need to express the densities $\rho^{(2)}$ and $\rho^{(3)}$ in this equation as a function of $\rho^{(1)}$. DDFT provides a strategy by mapping each state of the system instantaneously to a corresponding equilibrium situation \cite{marconi1999dynamic,marconi2000dynamic,archer2004dynamical, lowen2010density}.

For this purpose, we recall that an external potential enters the dynamical equation via (\ref{eq_U}). At each moment in time, DDFT assumes that the instant state of the system is caused by an effective external potential $\Phi_{\mathrm{ext}}$. This $\Phi_{\mathrm{ext}}$ intermittently takes the place of our physical external potential $u_{\mathrm{ext}}$. 

In equilibrium, density functional theory (DFT) implies that $\Phi_{\mathrm{ext}}$ is uniquely determined by the density $\rho^{(1)}$ \cite{hansen1990theory,singh1991density,evans1992density,marconi1999dynamic, marconi2000dynamic, archer2004dynamical,evans2010density,lowen2010density}. 
It follows by minimizing the grand canonical potential functional $\Omega$ 
\begin{equation}
\label{Omega}
\Omega\left[\rho^{(1)}\right]
=
\mathcal{F}_\mathrm{id}\left[\rho^{(1)}\right] +\mathcal{F}_\mathrm{exc}\left[\rho^{(1)}\right] 
+\mathcal{F}_\mathrm{ext}\left[\rho^{(1)}\right]
\end{equation}
with respect to $\rho^{(1)}$. Here,
\begin{eqnarray}
\mathcal{F}_\mathrm{id}\left[\rho^{(1)}\right]
&=&
k_BT\int \mathrm{d} \vect{X}\,\rho^{(1)}(\vect{X}) \left(\ln\left(\lambda^3\rho^{(1)}(\vect{X})\right)-1\right)\qquad
\end{eqnarray}
is the entropic free-energy functional for ideal non-interacting particles, with $\lambda$ the thermal de Broglie wave length \cite{wittkowski2010derivation}. 
An exact expression for the excess free-energy functional $\mathcal{F}_\mathrm{exc}[\rho^{(1)}]$, which contains all particle interactions beyond the idealized non-interacting limit, is typically not known and needs to be approximated. The third functional
\begin{equation}
\mathcal{F}_\mathrm{ext}\left[\rho^{(1)}\right] = \int \mathrm{d}\vect{X}\,\Phi_\mathrm{ext}(\vect{X})\rho^{(1)}(\vect{X}),  
\end{equation}
describes the interactions with the external potential, where the effect of a chemical potential is implicitly included into $\Phi_\mathrm{ext}$. 
Minimizing $\Omega$ with respect to $\rho^{(1)}$ leads to the equilibrium relation
\begin{equation}
\Phi_\mathrm{ext}(\vect{X})\,=
-k_BT\,\ln\left(\lambda^3\rho^{(1)}(\vect{X})\right)
-\left.\frac{\delta {\mathcal F}_\mathrm{exc}}{\delta \rho^{(1)}(\vect{X})}\right..
\label{rhoeq_equilibrium_prev}
\end{equation}

In equilibrium the swimmers are inactive ($f=0$). Then, we may further argue that the corresponding $N$-swimmer probability density $P^{\mathrm{eq}}$ solely depends on the overall potential $U=U(\vect{X}_1,\dots,\vect{X}_N)$ as in (\ref{eq_U}), but with $\Phi_{\mathrm{ext}}$ taking the place of $u_{\mathrm{ext}}$. Then, $P^{\mathrm{eq}}$ should follow the Boltzmann form
\begin{equation} 
P^{\mathrm{eq}} \propto \exp \left( -\beta U  \right),
\end{equation}
with $\beta=(k_\mathrm{B} T)^{-1}$.
Applying to this relation the positional gradient for the $i$th swimmer, we obtain
\begin{equation}
 \nabla_{\vect{r}_i} P^{\mathrm{eq}} = {}-\beta\, P^{\mathrm{eq}} \left( \nabla_{\vect{r}_i} \Phi_\mathrm{ext}(\vect{r}_i) +
 \nabla_{\vect{r}_i} \sum_{\substack{k\neq i}}^N u(\vect{r}_k,\vect{r}_i) \right).
\label{ygb_basis}
\end{equation}
We then follow (\ref{nbody}) and integrate out all coordinates from this relation except for those of the $i$th swimmer. Since all swimmers are identical, this leads to the so-called YGB relations of first order \cite{Gray_Gubbins, hansen1990theory}, 
\begin{equation}
 k_\mathrm{B} T \; \nabla_\vect{r} \rh{1}(\vect{X}) = {}- \rh{1}(\vect{X}) \nabla_\vect{r} \Phi_\mathrm{ext}(\vect{X}) - \int \mathrm{d}\vect{X}' \rh{2}(\vect{X},\vect{X}') \nabla_\vect{r} u(\vect{r},\vect{r}').
\end{equation}
The YGB relations of second order are obtained by integrating out from (\ref{ygb_basis}) all coordinates but those of the $i$th and one other swimmer \cite{Gray_Gubbins, hansen1990theory}, resulting in
\begin{eqnarray}
 k_\mathrm{B} T \;  \nabla_{\vect{r}'} \rh{2}(\vect{X},\vect{X}') &= &{}- \rh{2}(\vect{X},\vect{X}') \nabla_{\vect{r}'} \Phi_\mathrm{ext}(\vect{X}') -\rh{2}(\vect{X},\vect{X}') \nabla_{\vect{r}'} u(\vect{r},\vect{r}') \nonumber \\ &&- \int \mathrm{d}\vect{X}'' \rh{3}(\vect{X},\vect{X}',\vect{X}'') \nabla_{\vect{r}'} u(\vect{r}',\vect{r}'').
\end{eqnarray}
We then eliminate $\Phi_\mathrm{ext}$ from the last two equations by inserting (\ref{rhoeq_equilibrium_prev}). The resulting relations 
\begin{equation}
\int \mathrm{d}\vect{X}'\, \rho^{(2)}(\vect{X},\vect{X}')\nabla_{\vect r}u(\vect r,\vect r')
=
\rho^{(1)}(\vect{X})\nabla_{\vect r}\frac{\delta {\mathcal F}_\mathrm{exc}}{\delta\rho^{(1)}(\vect{X})}
\label{YBG1}
\end{equation} 
and
\begin{eqnarray}
\fl\int \mathrm{d}\vect{X}''\,\rho^{(3)}(\vect{X},\vect{X}',\vect{X}'') \nabla_{\vect r'}u(\vect r',\vect r'') &=& 
{}- k_\mathrm{B} T \;   \nabla_{\vect r'}\rho^{(2)}(\vect{X},\vect{X}') 
- \rho^{(2)}(\vect{X},\vect{X}')\nabla_{\vect r'}u(\vect r,\vect r')
\nonumber\\ 
&&{}
+k_\mathrm{B} T \rho^{(2)}(\vect{X},\vect{X}')  \;  \nabla_{\vect r'} \ln\left(\lambda^3\rho^{(1)}(\vect{X}',t)\right)  \nonumber \\
&&{}
+\rho^{(2)}(\vect{X},\vect{X}') \nabla_{\vect r'}\ddfrac{\delta {\mathcal F}_\mathrm{exc}}{\delta \rho^{(1)}(\vect{X}^\prime)} 
\label{YBG2}
\end{eqnarray} 
have the same structure as the corresponding ones in Ref.~\onlinecite{menzel2016dynamical}.

DDFT assumes that these relations are still instantly satisfied in non-equilibrium at each moment in time. All contained quantities are then assumed to be dynamical and non-equilibrium ones. In this way, they are inserted into the dynamical equation for $\rho^{(1)}$. Our assumption implies that the higher-order swimmer densities relax quickly when compared to the lower-order ones \cite{espanol2009derivation}. Since our motion at low Reynolds numbers is overdamped, it is conceivable that this adiabatic approximation leads to reasonable results. Previous comparison with particle simulations has confirmed this assertion qualitatively \cite{menzel2016dynamical}.

Altogether, we obtain from this procedure
\begin{equation}
\label{BBGKY1}
\frac{\partial\rho^{(1)}(\vect{X},t)}{\partial t} = -\nabla_{\vect{r}}\cdot(\vgr{\mathcal{ J}\!}_1+\vgr{\mathcal{ J}\!}_2+\vgr{\mathcal{ J}\!}_3)
- \nabla^\mathrm{or} \cdot(\vgr{\mathcal{ J}\!}_4+\vgr{\mathcal{ J}\!}_5+\vgr{\mathcal{ J}\!}_6),
\end{equation}
where $\curr{1},\dots,\curr{6}$ are \rev{current densities}. They are of similar structure as the corresponding quantities in Ref.~\onlinecite{menzel2016dynamical}, but particularly the active current densities $\curr{3}$ and $\curr{6}$ differ in the present case because of the transversal shift of the active force centers, see figure~\ref{fig:model}, 
\begin{eqnarray}
\label{DDFT_J1}
\fl \curr{1}
=
{}-\mu^{\mathrm{t}}\left( k_\mathrm{B} T \; \nabla_{\vect{r}}\rho^{(1)}(\vect{X},t)
+\rho^{(1)}(\vect{X},t)\nabla_{\vect{r}}u_\mathrm{ext}(\vect{r})
+\rho^{(1)}(\vect{X},t)\nabla_{\vect{r}}\frac{\delta {\mathcal F}_\mathrm{exc}}{\delta\rho^{(1)}(\vect{X},t)}\right)
\nonumber\\
{}-\int \mathrm{d} \vect{X}' { \vgr{\mu}^{\mathrm{tt}}_{\vect{r},\vect{r}'}}\cdot\Bigg(\rho^{(2)}(\vect{X},\vect{X}',t)
%\nonumber\\&&
\Bigg( k_\mathrm{B} T \; \nabla_{\vect{r}'}\ln\left(\lambda^3\rho^{(1)}(\vect{X}',t)\right) + \nabla_{\vect{r}'}u_\mathrm{ext}(\vect{r}') \nonumber \\
{}+\nabla_{\vect{r}'}\ddfrac{\delta {\mathcal F}_\mathrm{exc}}{\delta\rho^{(1)}(\vect{X}',t)}\Bigg)\Bigg),\\[.2cm]
\fl \curr{2}
=
{}-\int \mathrm{d}\vect{X}' \,\vgr{\mu}_{\vect{r},\vect{r}'}^{\mathrm{tr}}\,k_\mathrm{B} T \; \rev{\nabla^{\mathrm{or}^\prime}} \rho^{(2)}(\vect{X}, \vect{X}', t), \label{DDFT_J2}
\\[.2cm]
\fl \curr{3}
=
f\left({\vgr{\Lambda}^{\mathrm{tt}}_{\vect{r},\vect{r}}}\cdot\uvec{n}\rho^{(1)}(\vect{X},t)
+\int \mathrm{d}\vect{X'} {\vgr{\Lambda}^{\mathrm{tt}}_{\vect{r},\vect{r}'}}\cdot\uvec{n}'\rho^{(2)}(\vect{X}, \vect{X}', t)\right), \label{DDFT_J3}\\[.2cm] 
\fl \curr{4}
=
{}-\int \mathrm{d}\vect{X'} { \vgr{\mu}^{\mathrm{rt}}_{\vect{r},\vect{r}'}}\,\Bigg(
\rho^{(2)}(\vect{X}, \vect{X}', t) 
\Bigg(k_\mathrm{B} T \; \nabla_{\vect{r}'}\ln\left(\lambda^3\rho^{(1)}(\vect{X}',t)\right)  \nonumber \\
{}+\nabla_{\vect{r}'}u_\mathrm{ext}(\vect{r}') 
+\nabla_{\vect{r}'}\ddfrac{\delta {\mathcal F}_\mathrm{exc}}{\delta\rho^{(1)}(\vect{X}',t)}\Bigg)\Bigg), \label{DDFT_J4}\\[.2cm]
\fl \curr{5}
=
{}-\mu^{\mathrm{r}} k_\mathrm{B} T \; \nabla^\mathrm{or} \rho^{(1)}(\vect{X},t)
-\int \mathrm{d}\vect{X'} \,\vgr{\mu}^{\mathrm{rr}}_{\vect{r},\vect{r}'}\cdot k_\mathrm{B} T \; \rev{\nabla^{\mathrm{or}^\prime}} \rho^{(2)}(\vect{X}, \vect{X}', t), \label{DDFT_J5}\\[.2cm] 
\fl \curr{6}=
f\left({\vgr{\Lambda}^{\mathrm{rt}}_{\vect{r},\vect{r}}}\uvec{n}\rho^{(1)}(\vect{X},t)
+
\int \mathrm{d}\vect{X'} \,{\vgr{\Lambda}^{\mathrm{rt}}_{\vect{r},\vect{r}'}}\,\uvec{n}'\rho^{(2)}(\vect{X}, \vect{X}', t) \right).
\label{DDFT_J6}
\end{eqnarray}
We note that, in our case, $\vgr{\mathcal{ J}\!}_2=\mathbf{0}$ in (\ref{DDFT_J2}), and also the integral containing $\rho^{(2)}$ in (\ref{DDFT_J5}) vanishes. The reason is the spherical shape of our passive swimmer bodies, resulting in hydrodynamic interactions that do not depend on the swimmer orientations. 

For the excess functional, we choose a mean-field approximation 
\rev{
\begin{equation} 
{\mathcal F}_\mathrm{exc}=\frac{1}{2}\int \mathrm{d} \vect{X} \,\mathrm{d}\vect{X}' \rho^{(1)}(\vect{X}, t)\rho^{(1)}(\vect{X}', t)u(\vect{r},\vect{r}'),
\label{Fexc}
\end{equation} 
}
which is reasonable in our case of soft GEM-4 steric interaction potentials. 
Still, some pair densities $\rho^{(2)}$ remain in (\ref{DDFT_J1})--(\ref{DDFT_J6}). They are expressed in terms of $\rho^{(1)}$ using a 
dilute-limit Onsager-like approximation \cite{onsager1949effects}
\rev{
\begin{equation}
\rh{2}(\vect{X},\vect{X}',t)=\rh{1}(\vect{X},t)\rh{1}(\vect{X}',t) \exp\left(-\beta V(\vect{r},\vect{r}^{\prime})\right) .
\end{equation}
Here, $V(\vect{r},\vect{r}')=u(\vect{r},\vect{r}')$, if $\vect{r}\neq \vect{r}'$. For $\vect{r}-\vect{r}' \to \vect{0}$, we let $\beta V \to \infty$ to avoid the hydrodynamic divergence that appears in the unphysical situation of two swimmers being located at the same position. 
This corresponds to $\exp(-\beta \epsilon_0) \to 0$, which in our typical choice of parameters will be $\exp(-10)$ already without setting $\exp(-\beta V)$ to zero at $\vect{r}=\vect{r}'$.}
In this way, our dynamical equation for $\rh{1}$ is derived and finally closed. 

To demonstrate the power of our DDFT for circle swimmers, we now address the confinement in a spherically symmetric trap.
In particular, we focus on the effect of an increasing curvature of the swimming paths.

\section{Circle swimmers in a spherically symmetric trap}
\label{sec:trap}

To address planar geometries, we confine the center of mass of each swimmer $i$ as well as its two orientation vectors $\uvec{n}_i$ and $\uvec{u}_i$ to the Cartesian $xy$ plane so that $\uvec{n}_i \times \uvec{u}_i=\uvec{z}$. Then, one angle $\varphi_i$ is sufficient to parameterize the swimmer orientation, see our remarks below (\ref{torque}). We measure $\varphi_i$ relatively to the \rev{$x$ axis}.
Still, three-dimensional hydrodynamic interactions apply.
One possible realization of this geometry are swimmers confined to the interface between two immiscible fluids of identical viscosity.

Next, we specify the spherically symmetric confining external potential in (\ref{eq_U}). As in Ref.~\onlinecite{menzel2016dynamical}, we use a quartic potential
\begin{equation}
	u_\mathrm{ext}(\vect{r}_k)=V_0 \left( \frac{r_k}{\sigma} \right)^{\!4},
\end{equation}
centered in the origin, where $r_k=|\vect{r}_k|$. This potential is more shallow around the center and then shows a steeper increase than a harmonic trap, which partially emphasizes the effects that we address in the following. Yet the precise functional form is not relevant for their qualitative nature.

To evaluate our DDFT numerically, the finite-volume-method (FVM) partial-differential-equation solver FiPy \cite{Guyer_2009_CiSE} is employed. 
Our numerical grid \rev{is regular, quadratic in the $xy$ space, and} typically consists of $80\times80\times16$ grid points for the $x$, $y$, and $\varphi$ coordinates, respectively. 
(Non-orthogonal meshes might produce significant numerical errors due to the assumption of orthogonality by the solver \cite{pavelka,fipymanual}. We avoid this by using an orthogonal grid.) 

\rev{We only analyze the behavior in one single isolated trap. 
Nevertheless, numeric periodic boundary conditions are imposed in all directions for technical reasons to allow for Fast Fourier Transformation. To avoid unphysical feedback between particles through the walls of the box, long-ranged hydrodynamic interactions are cut at distances larger than half a box length.
Care is taken that the extension of the density cloud, before it basically decays to zero due to the external potential, is smaller than half a box length. In this way, the density cloud does not interact with itself through the periodic box boundaries.} However, the box is large enough to account for all hydrodynamic interactions within the effective confinement by the spherical trap. \rev{The steric interactions in (\ref{eqGEM4}) do not need to be cut as they quickly decay with increasing distance. If, instead of one single isolated trap, an array of periodically placed traps were to be regarded, one would have to account for the long-ranged hydrodynamic interactions between the individual traps including the periodic images of the system, e.g., via Ewald summation techniques \cite{ewald1921berechnung,beenakker1986ewald,brady1988dynamic}.} 

To display our results, we extract the spatial swimmer density
\begin{equation}
	\rho(\vect{r},t)=\int \mathrm{d} \varphi \; \rh{1}(\vect{r},\varphi,t)
\end{equation}
and the orientational vector field
\begin{equation}
	\left< \uvec{n} \right> (\vect{r},t) = \int \mathrm{d} \varphi \; \uvec{n}(\varphi) \; \rh{1}(\vect{r},\varphi,t)
\end{equation}
from our calculations. 
These two fields are indicated by color plots and by white arrows, respectively, in the figures referred to below. 
In these plots, the spatial density $\rho(\vect{r},t)$ is normalized by the density $\bar{\rho}$ averaged over the whole simulation box.

\begin{figure}
 \centering
\includegraphics[width=0.75\linewidth]{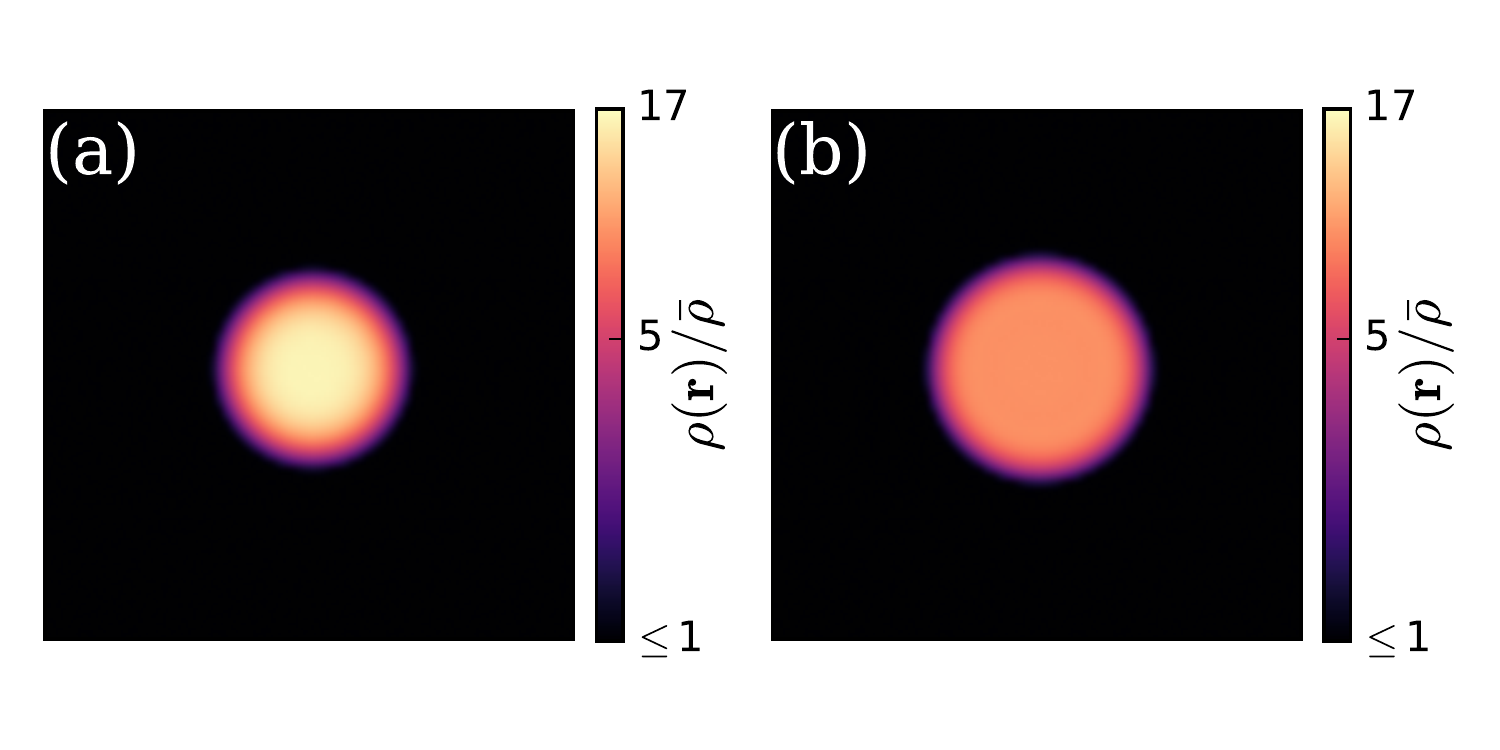}
\caption{Equilibrated initial state for the numerical evaluation of our DDFT for microswimmer suspensions. In this state, the active forces are still switched off, $f=0$. (a) Steric interactions not included, $\epsilon_0=0$. (b) Steric interactions included with strength $\epsilon_0=10 k_\mathrm{B}T$. Brighter color reflects higher density as given by the scale bars\rev{, where we used a logarithmic scale for illustration}. The local density $\rho(\vect{r})$ is normalized by the mean density $\bar{\rho}=2.78\times10^{-2}\sigma^{-2}$ in the whole simulation box. }
\label{fig:equilibrium}
\end{figure}

As an initial condition, we start from randomized density distributions. The system is then equilibrated in the trap with self-propulsion switched off, $f=0$. The density quickly relaxes into a radially decaying distribution with a small central dip stemming from steric repulsion, see figure~\ref{fig:equilibrium}. 
We measure time $t$ in units of $\sigma^2 / (\mu_t k_\mathrm{B} T)$. 
At $t=0$, self-propulsion is switched on. 
Such a process could be achieved in reality, for instance, using light-activated synthetic swimmers \cite{volpe2011microswimmers,buttinoni2012active,palacci2013living, buttinoni2013dynamical,hagen2014gravitaxis,samin2015self}. 
If, for example, activation of self-propulsion is sensitive to the wavelength of the irradiated light \cite{palacci2013living},  confinement might be achieved simultaneously by optical trapping using light of a different frequency. 

To characterize the relative strength of self-propulsion, often the dimensionless P\'{e}clet number $\mathrm{Pe}_\mathrm{tr}$ is introduced \cite{hennes2014self}. In our context, it measures the ratio of active to diffusive passive motion on a relevant length scale, here set by $\sigma$. Therefore, 
\begin{equation}
	\mathrm{Pe}_\mathrm{tr}=\frac{v_\mathrm{s} \sigma}{\mu^{\mathrm{t}} k_\mathrm{B} T}.
\end{equation} 
In the following, we concentrate on microswimmers of significant activity, $\mathrm{Pe}\gg 1$. 
Moreover, we may in the case of circle swimming analogously define a rotational P\'{e}clet number, 
\begin{equation}
	\mathrm{Pe}_\mathrm{rot}=\frac{\omega_\mathrm{s}}{\mu^{\mathrm{r}} k_\mathrm{B} T} 
      =  \frac{4 a^2}{3 R_\mathrm{s} \sigma} \mathrm{Pe}_\mathrm{tr}.
\end{equation}
For $\mathrm{Pe}_\mathrm{rot} \approx 0$, the curvature of the swimmer trajectory is negligible. 
In our numerical scheme, we directly set the parameters determining the geometry of the swimmers in figure~\ref{fig:model}. The corresponding P\'{e}clet numbers can then be extracted by calculating $v_\mathrm{s}$ and $\omega_\mathrm{s}$ from (\ref{mobility}) and (\ref{Lambda_tt})--(\ref{defminus}) for $i=j$. 

It turns out that increasing the character of circle swimming, i.e., decreasing the radius of the unperturbed swimming path, see figure~\ref{fig:circle_radius}, has a qualitative effect on the appearance of the trapped swimmer suspension. To demonstrate this, we first further analyze some results of straight swimming \cite{menzel2016dynamical} obtained by our modified simulation scheme and then compare with the results for circle swimming.

\begin{figure*}
\centering
\includegraphics[width=\linewidth]{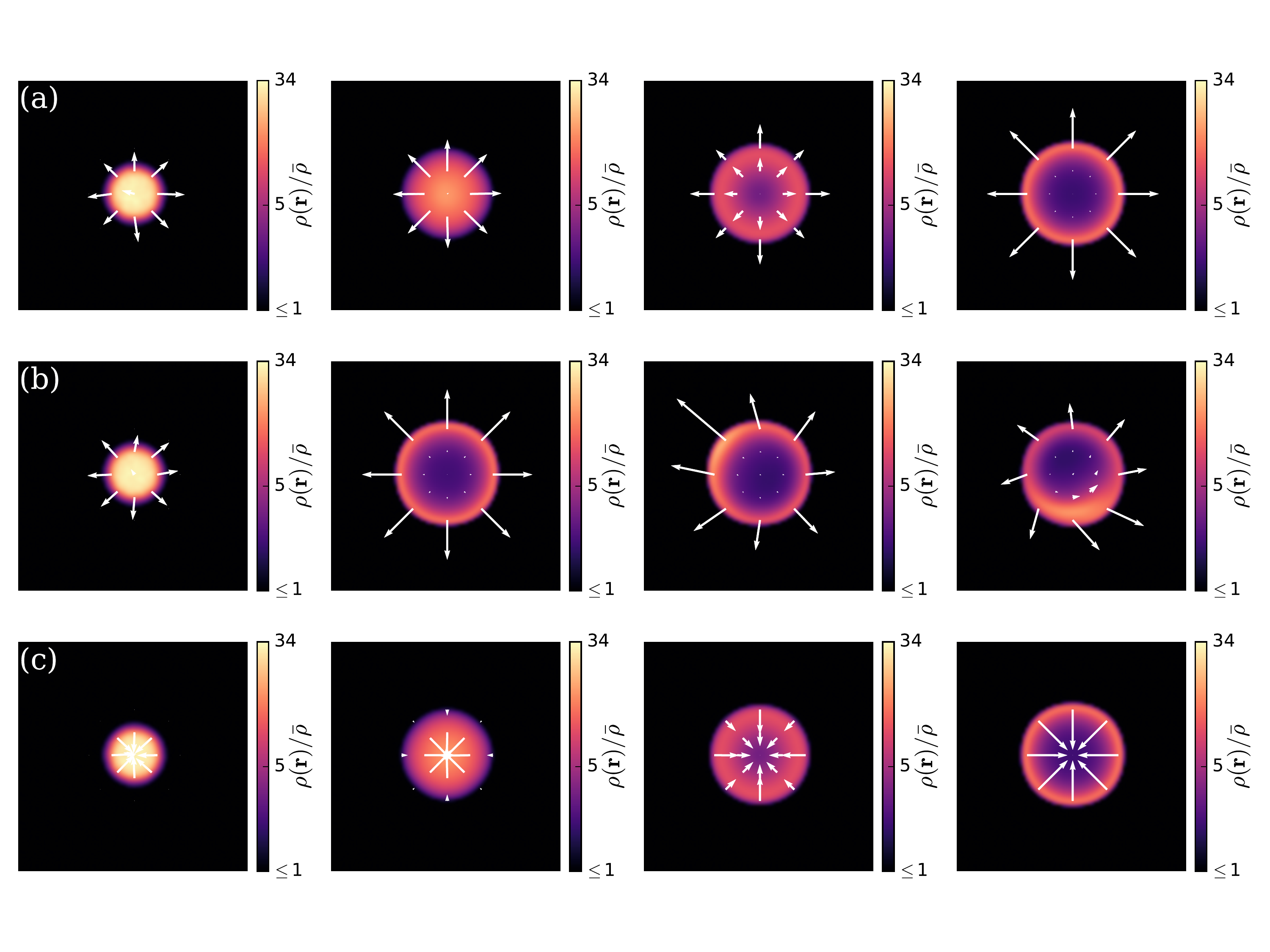}
\caption{
Each row represents a time series (from left to right) for a suspension of straight-propelling swimmers in a quartic external trapping potential for the spatial density $\rho(\mathbf{r})$ normalized by the mean density $\bar{\rho}=7.56\times10^{-3} \sigma^{-2}$ in the simulation box (color plots, brighter color indicating higher density\rev{, logarithmic scale}) and for the local swimmer orientation $\langle\uvec{n}\rangle(\vect{r})$ (white arrows). Our parameters are set to 
$a=0.25 \sigma$, 
$L=3a=0.75\sigma$,
$\alpha=0.4$,
$\gamma=0$,
$N=4$, 
$V_0=0.1 k_\mathrm{B}T$, 
and
$\epsilon_0=10 k_\mathrm{B}T$.
For each row, the system was pre-equilibrated with self-propulsion switched off, $f=0$. Then, self-propulsion is turned on at time $t=0$, with times measured in units of $\sigma^2 / (\mu_t k_\mathrm{B} T)$. 
(a) $f=250 k_\mathrm{B}T / \sigma>0$ for pushers, implying $\mathrm{Pe}_\mathrm{tr}\approx53$, with hydrodynamic interactions between the swimmers switched off. Snapshots provided for times $t=0.01, 0.05, 0.1, 0.3$, where the last image already shows the steady state. 
(b) Same as in (a), now including hydrodynamic interactions. The latter destabilize the high-density ring and lead to the formation of a high-density spot. Over time, for the chosen parameters, the self-propulsion directions in this spot lean towards one side by another spontaneous symmetry breaking. The spot then starts to move along the ring contour, smearing out in the process. Times: $t=0.01,0.5,1.4,1.8$. 
(c) Same as in (b), but for pullers $f=-250 k_\mathrm{B}T / \sigma<0$, which restricts the spot formation. Times: $t=0.01, 0.05, 0.1, 0.3$. Inverting $f$ makes the swimmers propel into the opposite direction $-\uvec{n}$, see figure~\ref{fig:model}, which makes the white arrows point inward in the case of pullers. As in (a), the last picture shows the steady state of the system.
}
\label{fig:straight}
\end{figure*}

\subsection{Straight swimming}

Straight motion of the individual swimmers is enforced in our approach by setting $\gamma=0$, see figure~\ref{fig:model} \cite{menzel2016dynamical}. For straight propelling objects under spherical confinement, the formation of a high-density ring has been reported several times \cite{nash2010run,hennes2014self,menzel2015focusing, yan2015force, menzel2016dynamical}. In agreement with previous studies, the formation of a high-density ring can be reproduced after switching on the active drive in our simulations. This ring is particularly symmetric when we switch off the hydrodynamic interactions between the swimmers, see figure~\ref{fig:straight}~(a). Its approximate radius is determined by balancing the active forward drive of the swimmers with the confining external potential force, leading to $R_\mathrm{ring}\sim\sigma( {v_\mathrm{s} \sigma}/{4 \mu^{\mathrm{t}} V_0} )^{1/3}
=\sigma( {\mathrm{Pe}_\mathrm{tr}  k_\mathrm{B} T}/{4 V_0} )^{1/3}$.

In the next two rows, figure~\ref{fig:straight} shows the behavior when hydrodynamic interactions between the swimmers are included as prescribed by our DDFT. They have a qualitative impact. The high-density ring at the investigated propulsion strengths develops a tangential instability and the circular symmetry is broken. Also this effect has been described before \cite{nash2010run,hennes2014self,menzel2016dynamical}. The swimmers tend to polarly order in the emerging high-density spot while propagating against the confining potential. Consequently, they collectively pump the surrounding fluid into the opposite direction. Thus the configuration was referred to as a ``hydrodynamic fluid pump'' \cite{hennes2014self,menzel2016dynamical}. Here, we observe that the effect is stronger in figure~\ref{fig:straight}~(b), which depicts the result for pushers, $f>0$. In contrast to that, figure~\ref{fig:straight}~(c) was obtained with the sign of the active forces flipped to $f<0$, describing a suspension of pullers, yet with all other parameters unchanged. Obviously, the tangential symmetry breaking is restricted in the latter case. 

The cause of this spontaneous symmetry breaking was attributed in Ref.~\cite{hennes2014self} to a positive feedback mechanism. If a spot of higher density appears on the ring, with the swimmers collectively pushing against the external potential, the resulting oppositely oriented fluid flow rotates nearby swimmers towards the high-density area. Consequently, they join the spot of higher concentration. In our DDFT, this effect is included by the contribution $\sim\bm{\mu}^{\mathrm{rt}}_{\mathbf{r},\mathbf{r}'}\nabla_{\mathbf{r}'}u_{\mathrm{ext}}(\mathbf{r}')$ to the \rev{current density} $\curr{4}$ in (\ref{DDFT_J4}). 
Additionally, pushers actively generate inward flows from their sides, see figure~\ref{fig:model}. When the swimmers are pointing outward on the ring, this further supports their lateral concentration, see the illustration in figure~\ref{fig:pushervspuller}~(a). Here, these active contributions are represented by the second term in the current density $\curr{3}$ in (\ref{DDFT_J3}). In contrast to that, for pullers, the actively induced flow fields are inverted. This in effect repels outward pointing swimmers on the ring from each other, see also our schematic illustration in figure~\ref{fig:pushervspuller}~(b).
The qualitative schematics in figure~\ref{fig:pushervspuller}~(c)--(f) indicate that also the curvature of the high-density ring may have a significant impact via the \rev{current density} $\curr{6}$ and lead to differences between pushers and pullers. 
\begin{figure}
\centering
 \includegraphics[width=\linewidth]{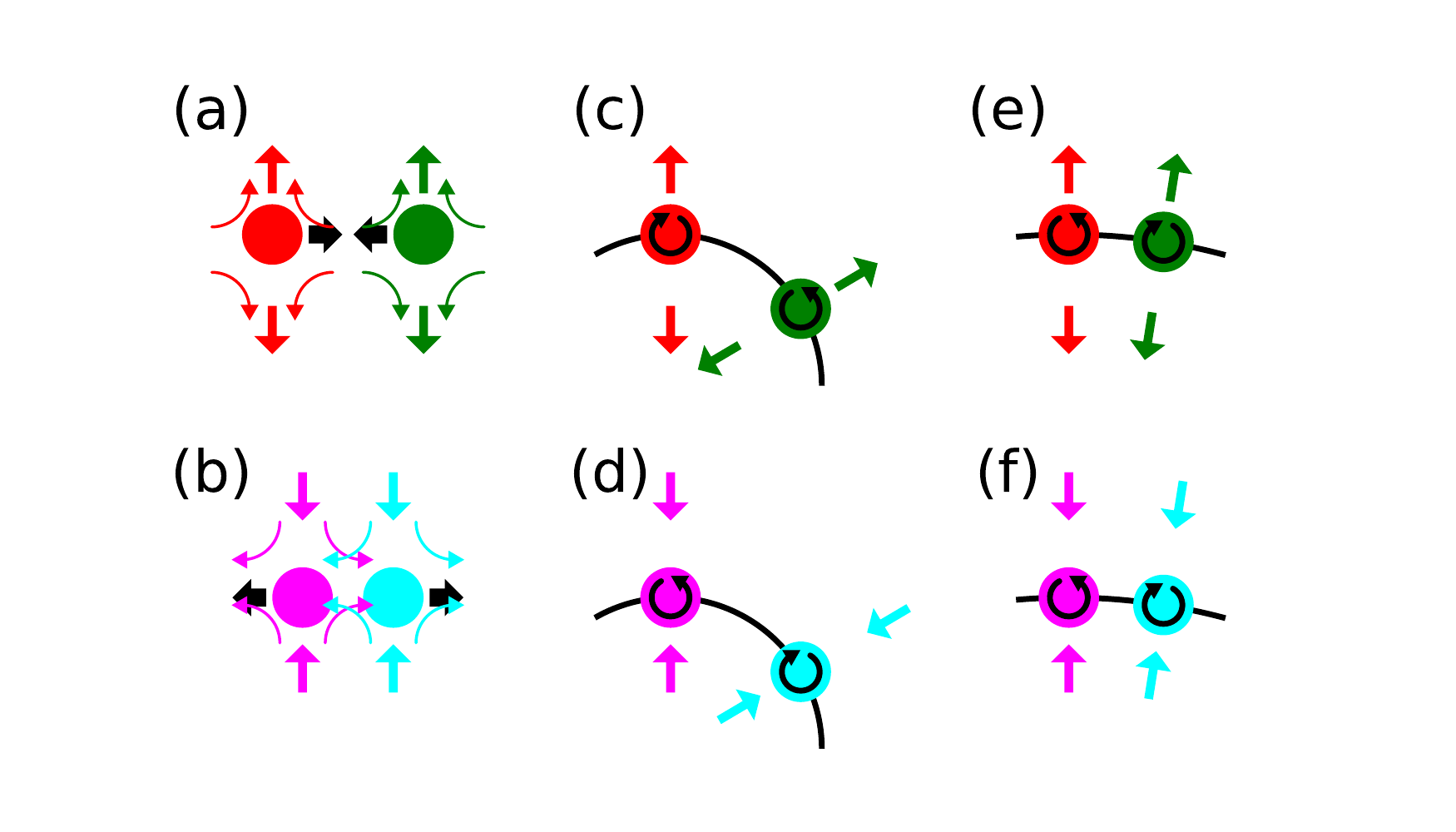}
\caption{
Effects that can contribute to the observed varying tendency of forming a high-density spot for pushers and pullers in figure~\ref{fig:straight}~(b) and (c), respectively.
Large straight arrows of the same color as the swimmer bodies represent the active forces, while the
smaller curved arrows indicate the corresponding flow fields.
(a) Due to their actively induced transversal inflow of fluid, see figure~\ref{fig:model}, pushers laterally attract each other hydrodynamically. This supports spot formation in figure~\ref{fig:straight}~(b) when the swimmers are aligned next to each other on a high-density ring. (b) In contrast to that, pullers laterally repel each other hydrodynamically, which counteracts a spot formation, see figure~\ref{fig:straight}~(c). Both effects are described by the contribution $\sim\vgr{\Lambda}^{\mathrm{tt}}_{\vect{r},\vect{r}'}$ to the \rev{current density} $\curr{3}$ in (\ref{DDFT_J3}). 
(c)--(f) Different effects are possible for the active rotation--translation coupling described by the contribution $\sim\vgr{\Lambda}^{\mathrm{rt}}_{\vect{r},\vect{r}'}$ to the \rev{current density} $\curr{6}$ in (\ref{DDFT_J6}). 
(c) For pushers next to each other on a high-density ring of high curvature, the inward pointing force center of one swimmer is in close vicinity of the body of the other swimmer, and vice versa. This leads to actively induced rotations of the swimmers and their propulsion directions towards each other, supporting the formation of a high-density spot. 
(d) The situation is reversed for pullers, leading to effective rotations away from each other.
(e) In contrast to the configuration in (c), for low curvature of the high-density ring, the outward pointing active force of one swimmer is closer to the body of the other swimmer, and vice versa. In this way, the swimmers tend to turn away from each other. 
(f) Along the same lines, pullers also for low curvature of the high-density ring turn away from each other, again counteracting the formation of a high-density spot.
}
\label{fig:pushervspuller}
\end{figure}
The relative magnitudes of all these different contributions basically involve all system parameters, i.e., temperature $T$, the viscosity $\eta$ of the fluid, the strength of the active force $f$, the nature of the swimmer (pusher vs.\ puller), the strength and radius of the steric interactions, and the strength of the confinement.

After the formation of the high-density spot, see figure~\ref{fig:straight}~(b), at strong enough active drive $f>0$, we observe yet another spontaneous symmetry breaking. 
In the rightmost snapshot of figure~\ref{fig:straight}~(b), the averaged self-propulsion directions do not point radially outward any more. Instead, they have tilted to one side towards the tangent of the previous high-density ring. For straight swimming objects, the selection of one of the two tilting directions depends solely on small variations in the initialization of the system. 

As a result of the tilting, the high-density spot starts to circle around the trap, smearing out the faded ring to some extent. Depending on the parameters, we may nearly recover a high-density ring, however, now with the swimmer orientations \textit{not} pointing outward.
Interestingly, for suspensions of pullers at elevated $|f|$, we so far have not observed this behavior. Instead, again a ring of radially oriented swimmers emerges, see figure~\ref{fig:straight}~(c). It appears approximately in the same way as for the case without hydrodynamic interactions in figure~\ref{fig:straight}~(b). This behavior is in line with our interpretation of the role of the current density $\curr{3}\sim f$ of repelling pullers from each other.

We note that the active \rev{current density} $\curr{6}$ in (\ref{DDFT_J6}) has the potential to drive the secondary spontaneous symmetry breaking observed in the rightmost snapshot of figure~\ref{fig:straight}~(b). Comparing the strength of $\curr{6}$ with the one of $\curr{4}$ may also explain the initial formation of the high-density spot as a first instability and then the observed secondary instability. First, on the high-density ring, the swimmers on average feature a larger mutual separation, see the second snapshot of figure~\ref{fig:straight}~(b). Then, at these larger interswimmer distances $r_{ij}$, the contribution in the \rev{current density} $\curr{4}$ driving the spot formation scales as $\sim |\vgr{\mu}^{\mathrm{rt}}_{ij}|\sim r^{-2}_{ij}$. In contrast to that, in the active current density $\curr{6}$, we find a scaling $\sim r^{-3}_{ij}$ at large interswimmer distances (the two oppositely oriented active forces of each swimmer together appear as a force dipole at larger distances, which reduces the exponent in the scaling of $\vgr{\Lambda}^{\mathrm{rt}}_{ij}$ to $\sim r^{-3}_{ij}$). Therefore $\curr{4}$ dominates and can drive the spot formation.
At reduced separation in the high-density spot, the active forces are resolved individually and the influence of $\curr{6}$ can become substantial when comparing with $\curr{4}$. Now both scale as $\sim r^{-2}_{ij}$, but for elevated $|f|$ the importance of $\curr{6}$ grows.

\begin{figure*}
 \centering
\includegraphics[width=\linewidth]{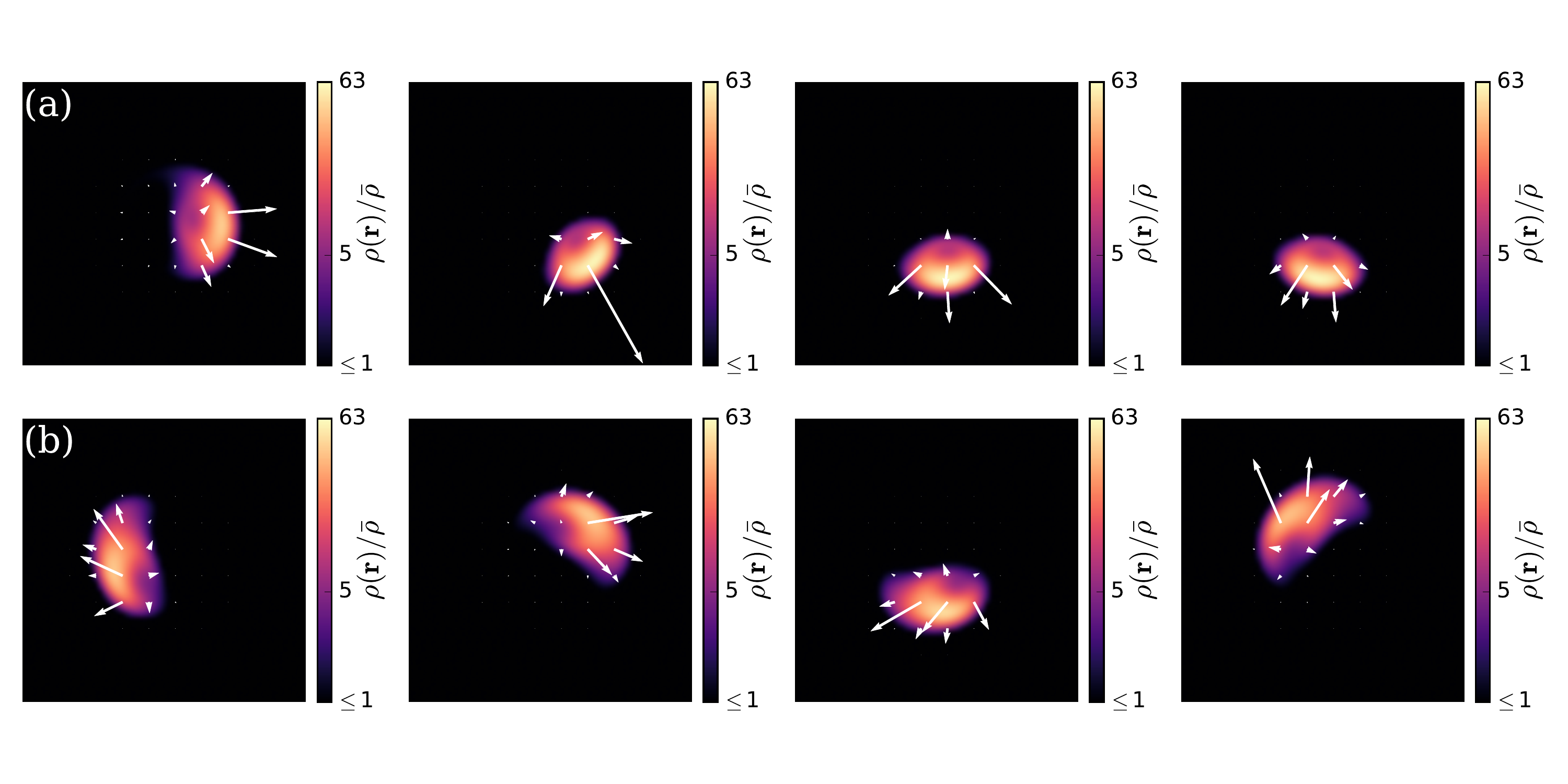}
\caption{Influence of a (small) biaxiality, quantified by the parameter $\gamma$, on the motion of the high-density spot formed by pushers ($f>0$). The type of presentation is identical to figure~\ref{fig:straight}, but the parameters are given by 
$a=0.5 \sigma$, 
$L=8a=4.\sigma$,
$\alpha=0.4$,
$N=4$, 
$V_0=1. k_\mathrm{B}T$,
$\epsilon_0=5 k_\mathrm{B}T$, and
$f=100 k_\mathrm{B}T/\sigma$. In the plot, the local density $\rho(\vect{r})$ is normalized by \rev{the} mean density $\bar{\rho}=6.47\times10^{-2} \sigma^{-2}$ in the whole simulation box.
Self-propulsion is switched on at $t=0$, the snapshots in each row are taken at $t=2,2.3,2.6,2.9$. 
(a) For straight-swimming pushers ($\gamma=0, \mathrm{Pe}_\mathrm{tr}=23$) the high-density spot only slowly migrates around the trap, the sense of motion resulting from spontaneous symmetry breaking.
(b) For weak circle swimmers ($\gamma=0.01,\mathrm{Pe}_\mathrm{tr}=23,\mathrm{Pe}_\mathrm{rot}=0.61$) the high-density spot persistently moves around the trap, with the sense of motion affected by the sense of circle swimming. 
}
\label{fig:spots_snaps}
\end{figure*}

\subsection{Circle swimming}

We now turn to increasingly biaxial swimmers for $\gamma\neq0$, see figure~\ref{fig:model}. Depending on the values of both parameters $\alpha$ and $\gamma$, the unperturbed individual swimmers then show circular trajectories, see 
figure~\ref{fig:circle_radius}.
Particularly, we analyze the changes in the behavior of the suspension when we stepwise increment $\gamma$. For each value of $\gamma$, we again start from an equilibrated passive initial system and then switch on self-propulsion at $t=0$, as before. 

By and large, we do not observe abrupt modifications in the overall behavior. Instead it changes rather gradually with increasing $\gamma$. For small $\gamma \neq 0$, the behavior of the straight swimming objects is reproduced qualitatively. Only for pushers of stronger active drive $f>0$, we note an illustrative alteration. 
While the sense of circling of the high-density spot around the trap as illustrated in the rightmost snapshot of figure~\ref{fig:straight}~(b) resulted from spontaneous symmetry breaking for $\gamma\neq0$ and could be clockwise or counterclockwise, it is now increasingly dictated by the sense of the circular swimming trajectory. A comparison between straight swimmers and weak circle swimmers is included by figure~\ref{fig:spots_snaps}.

Remarkably, the overall appearance of the suspension changes qualitatively when the nature of circle swimming becomes more pronounced. In our set of parameters we achieve this by increasing $\gamma$. The bending of the swimmer trajectories has a \textit{localizing} effect, 
as illustrated in figure~\ref{fig:circle_steady_states}. There, all snapshots show the long-term behavior of the corresponding suspension. From left to right in each row, the strength of circle swimming grows. 
Due to their persistent self-rotation, the outward propagation of the swimmers against the confining trapping potential is restricted. As a consequence, the concentration of the swimmers in the center of the trap increases. At high enough $\gamma$, the density is again peaked around the center of the trap.
Comparing figure~\ref{fig:circle_steady_states}~(a), where hydrodynamic interactions have been switched off, to figure~\ref{fig:circle_steady_states}~(b) and (c), we infer that hydrodynamic interactions significantly delay the localization around the center of the trap with increasing $\gamma$. Yet, at high enough values of $\gamma$ (rightmost column in figure~\ref{fig:circle_steady_states}) the localization dominates in all cases.  
Comparing pushers and pullers in figure~\ref{fig:circle_steady_states}~(b) and (c), respectively, we note the more persistent nature of the high-density ring in the case of pullers at smaller values of $\gamma$, before the collapse towards the center of the trap occurs.

To quantify the modified appearance of the suspension with increasing $\gamma$, we introduce the following order parameters. First, we evaluate 
\begin{equation}
	K(t)=\frac{1}{N}\left|  \int \mathrm{d}\vect{r} \mathrm{d} \varphi \; \exp(\mathrm{i} \vartheta) \; \rh{1}(\vect{r},\varphi,t) \right|, 
\end{equation}
where in this expression spatial positions $\mathbf{r}$ are parameterized by polar coordinates $\mathbf{r}=(r,\vartheta)$.  
$K(t)$ becomes nonzero when a tangential instability occurs that breaks the circular symmetry of a high-density ring, leading to an off-center high-density spot. 

Next, we define
\begin{equation}
	M_\mathrm{r}(t)=\frac{1}{N} \int \mathrm{d} \vect{r} \mathrm{d} \varphi \; (\uvec{v}_\mathrm{s} \cdot \uvec{r}) \; \rh{1}(\vect{r},\varphi,t),
\end{equation}
with $\uvec{v}_\mathrm{s}$ for each swimmer denoting the hypothetical instantaneous unperturbed direction of self-propulsion. For $\gamma=0$, $\uvec{v}_\mathrm{s}$ points along $\pm \uvec{n}$ according to the sign of $f$, but it becomes slightly tilted towards $\uvec{u}$ for $\gamma \neq 0$.
$M_\mathrm{r}(t)$ quantifies the overall degree of swimmer orientations along the radial direction.

In analogy to that, to quantify the ordering of the swimmer orientations along one of the two tangential directions, the order parameter
\begin{equation}
	M_\mathrm{t}(t)=\frac{1}{N} \left| \int \mathrm{d} \vect{r} \mathrm{d} \varphi \; \left(\uvec{v}_\mathrm{s} \cdot (\uvec{r} \times \uvec{z}) \right) \; \rh{1}(\vect{r},\varphi,t) \right|
\end{equation}
is evaluated. 
In the absence of any local orientational order, both $M_\mathrm{r}(t)$ and $M_\mathrm{t}(t)$ vanish. 
For steady-state systems, 
all three of the above order parameters no longer depend on time in the long-term limit. 

\begin{figure*}
\centering
\includegraphics[width=\linewidth]{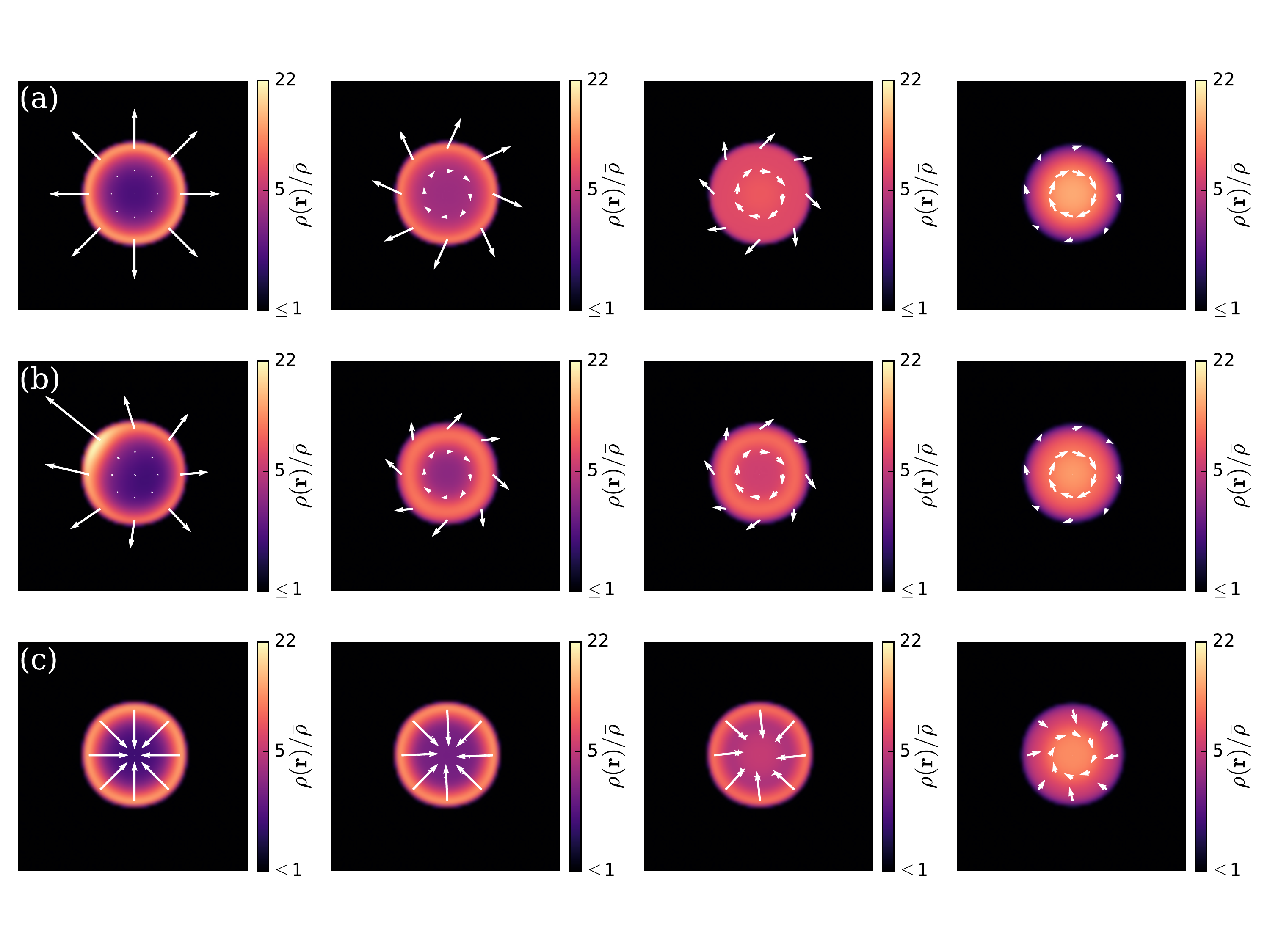}
\caption{Long-term behavior for different strengths of circle swimming, namely
$\gamma=0,0.01,0.02,0.04$ from left to right in each row. 
The other parameters \rev{and the type of presentation} are the same as in figure~\ref{fig:straight}. 
In all depicted cases, a \textit{localizing} effect of circle swimming becomes obvious. Pronounced circle swimming leads to concentration of the swimmers around the center of the trap, see the rightmost column. 
(a) Switching off hydrodynamic interactions supports the localization around the center of the trap. 
(b) For pushers ($f>0$), we observe with increasing $\gamma$ that first the high-density spot smears out to a high-density ring that broadens and for high $\gamma$ collapses towards the center. 
(c) For pullers ($f<0$) the high-density ring appears a bit more stable at lower values of $\gamma$, but again a concentration around the center occurs at high $\gamma$. 
}
\label{fig:circle_steady_states}
\end{figure*}

Figure~\ref{fig:mnm} shows \rev{the} long-term values of the order parameters $K$, $M_\mathrm{r}$, and $M_\mathrm{t}$ with increasing biaxiality and degree of circle swimming $\gamma$. 
\begin{figure}
\centering
 \includegraphics[width=0.75\linewidth]{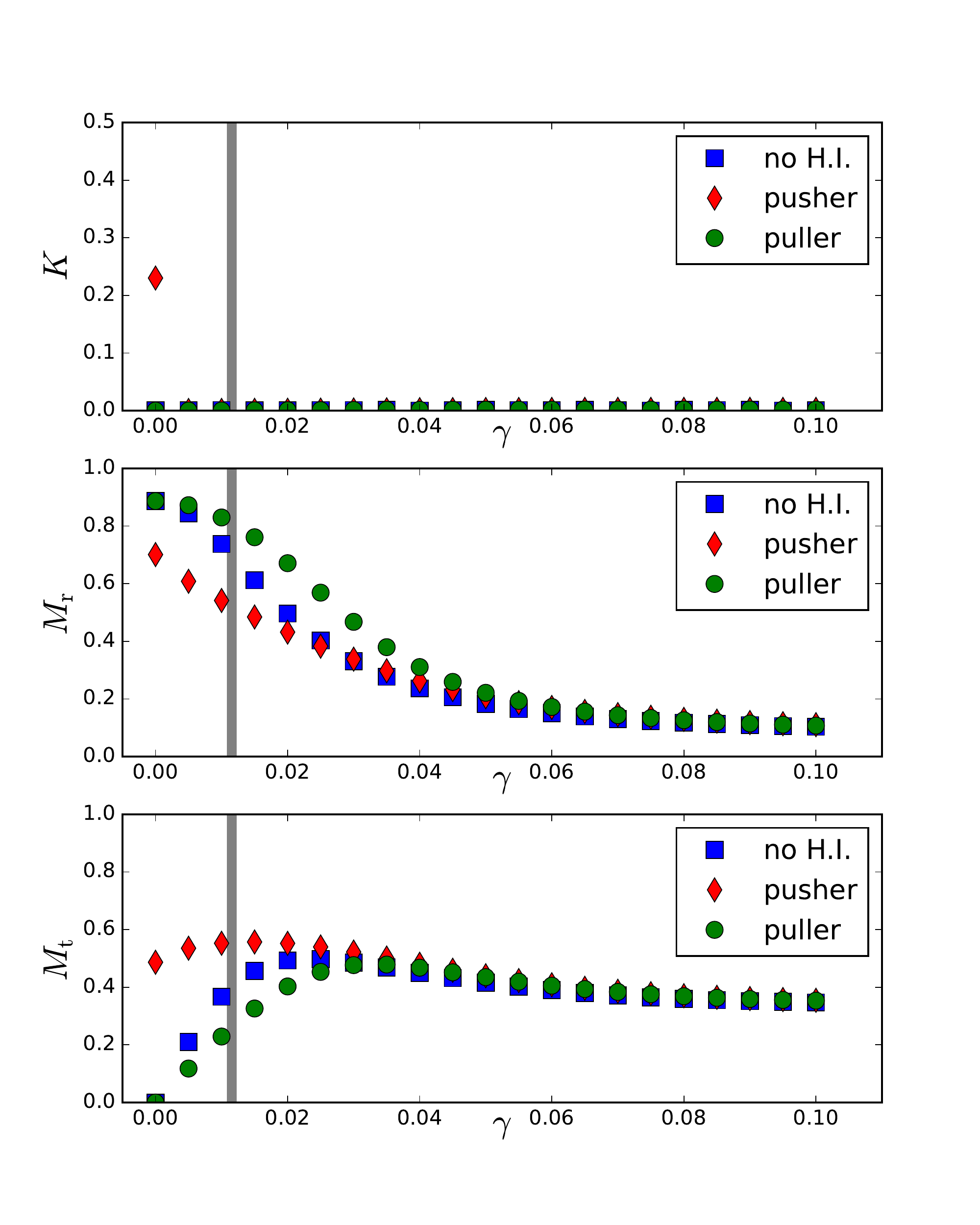}
\caption{
Order parameters $K$, measuring the degree of off-center density concentration in a high-density spot, as well as $M_\mathrm{r}$ and $M_\mathrm{t}$, measuring the degrees of swimmer orientations along the radial and one of the tangential directions, respectively, for the systems in figure~\ref{fig:circle_steady_states} 
with increasing biaxiality parameter $\gamma$. 
Again, the situation without hydrodynamic interactions between the swimmers (``no H.I.''), the case of $f>0$ (``pusher''), and the case of $f<0$ (``puller'') are depicted. 
Generally, with increasing $\gamma$, off-center concentration in non-rotationally symmetric structures diminishes (drop of $K$) and the swimmers tilt away from the radial direction (decreasing $M_\mathrm{r}$). 
We find smooth intermediate transitions for $M_\mathrm{r}$ and $M_\mathrm{t}$ around the value of $\gamma$ that leads to $R_\mathrm{s}=R_\mathrm{ring}$, as indicated by the vertical gray lines.
\rev{For the parameter chosen here, $K$ drops to zero at very low biaxiality, which is not true for all parameters, see, e.g., figure \ref{fig:spots_snaps}.}
}
\label{fig:mnm}
\end{figure}
When hydrodynamic interactions are switched off, for $\gamma=0$ a high-density ring is formed with the swimmers radially aligned, see figure~\ref{fig:straight}~(a). Therefore, $K$ and $M_\mathrm{t}$ are low, while $M_\mathrm{r}$ is high.

Including hydrodynamic interactions, pullers ($f<0$) here behave in a very similar way, see also figure~\ref{fig:circle_steady_states}~(c). 
In contrast to that, pushers ($f>0$) show a concentration in high-density spots for $\gamma=0$, see figures~\ref{fig:spots_snaps} and  \ref{fig:circle_steady_states}~(b), leading to an elevated value of $K$. Moreover, the self-propulsion directions in this high-density spot by spontaneous symmetry breaking can lean towards one of the two tangential directions, see figures~\ref{fig:straight}~(b) and \ref{fig:spots_snaps}. 
Therefore, $M_\mathrm{r}$ and $M_\mathrm{t}$ are reduced and elevated, respectively, when compared to the other systems in figure~\ref{fig:mnm}. 

As the degree of circle swimming increases with $\gamma$ and the swimmers tilt away from the radial outward direction, $M_\mathrm{r}$ generally decreases. 
$M_\mathrm{t}$ first increases as the orientational order shifts from radial to tangential. It then saturates and again slightly decays for high $\gamma$, i.e., for small swimming radii.
The latter slow decay is supported by the increasing localization in the center of the trap where orientational order vanishes by the overall rotational symmetry.
The smooth changes of $M_\mathrm{r}$ and $M_\mathrm{t}$ in figure~\ref{fig:mnm} indicate that the transition from off-center high-density rings or spots to centrally localized distributions with increasing $\gamma$ is rather continuous.
This transition should occur when the radius $R_\mathrm{s}$ of the unperturbed swimmer trajectories and the characteristic radius of the trap $R_\mathrm{ring}$ become approximately identical. We have indicated the corresponding value of $\gamma$ in figure~\ref{fig:mnm} by the vertical gray lines.

To also quantify the depletion of the swimmer density in the center of the trap when high-density rings or off-center high-density spots occur, in contrast to the central accumulation when the localizing effect of circle swimming becomes strong, we introduce additional order parameters
\begin{equation}
 O_\nu(t)= \int\mathrm{d}\vect{r}\mathrm{d}\varphi \; 
  J_\nu\left( r
  / R_\mathrm{ring}\right) \; \rh{1}(\vect{r},\varphi,t)
\end{equation}
for $\nu=0$ and $1$. Here, $J_\nu$ are the Bessel functions of first kind. 
By construction, $O_0(t)$ is large when the density is concentrated in the center of the trap, 
is elevated for off-center distributions.

As demonstrated by figure~\ref{fig:bessel}, 
\begin{figure}[t]
\centering
 \includegraphics[width=0.75\linewidth]{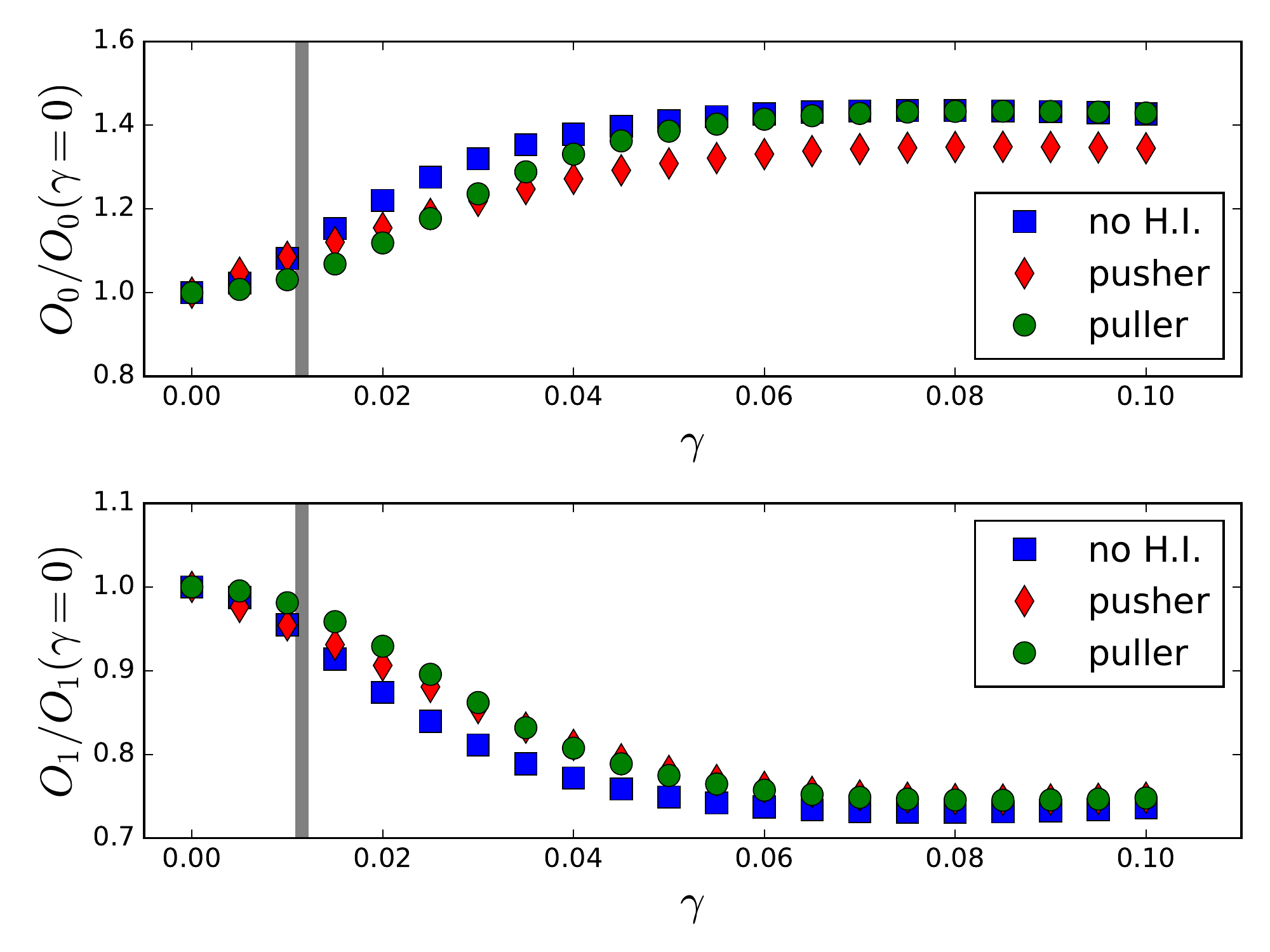}
\caption{Same as in figure~\ref{fig:mnm}, but for the order parameters $O_0$ and $O_1$ that measure the degree of near-center and off-center concentration, respectively. At smaller values of the biaxiality parameter $\gamma$, $O_0$ is low and $O_1$ is high when a high-density ring or a high-density off-center spot has formed. In contrast to that, high $O_0$ and low $O_1$ signal a localization of the density around the center of the trap for pronounced circle swimming at high values of $\gamma$ (see also the rightmost column in figure~\ref{fig:circle_steady_states}). 
Again, the vertical gray lines indicate the value of $\gamma$ implying $R_\mathrm{s}=R_\mathrm{ring}$. 
Apparently, hydrodynamic interactions slightly counteract the concentration around the center. For strong circle swimming, pullers appear more concentrated around the center than pushers. 
} 
\label{fig:bessel}
\end{figure}
the transitions as a function of the biaxiality parameter $\gamma$ are again smooth. Yet, the increasing localization in the center of the trap for increasing $\gamma$ is obvious. Particularly in the transitional regime, that is, for intermediate values of $\gamma$, hydrodynamic interactions apparently counteract localization in the center of the trap. Moreover, for our set of parameters and at large $\gamma$, the central concentration of pushers is slightly lower than the one for pullers, if only the sign of $f$ is inverted and all other parameters are kept the same.

\section{Conclusions}
\label{sec:Conclusions}

In summary, we have presented a microscopic statistical approach in the framework of dynamical density functional theory (DDFT) on active circle swimmers. Hardly any real microswimmer is a perfectly symmetric straight swimmer. Therefore, investigations on the effect of bent migration trajectories are mandatory. 

Our theory captures self-propulsion along swimming paths of different preferred curvature, steric and hydrodynamic interactions between the microswimmers, as well as confinement by an external potential. In contrast to many previous descriptions, the curved motion in our case is not directly imposed by an effective torque or angular frequency on the swimmer body. Here, it naturally follows from the geometric structure of our microscopic minimal swimmer model and resulting hydrodynamic effects. 

Persistently bent swimming trajectories reduce the global mobility of the swimmers. To study this \textit{localizing} effect, we analyzed the behavior of microswimmer suspensions in a circularly symmetric trapping potential for increasing degree of circle swimming. Moreover, we distinguished between pusher and puller circle swimmers, and also studied the effect of hydrodynamics by comparison with switched-off hydrodynamic interactions between the swimmers. 

Straight swimming objects tend to spread out towards the confinement until their active drive is balanced by the confining potential \cite{nash2010run,hennes2014self,menzel2015focusing,yan2015force, menzel2016dynamical}. This leads to high-density rings. Such rings may get unstable due to hydrodynamic interactions, particularly for pusher swimmers, leading to the formation of off-center high-density spots \cite{nash2010run,hennes2014self,menzel2016dynamical}. We have further investigated and quantified this scenario. 

Circle swimming can qualitatively affect the behavior. 
Increasing the degree of circular self-propulsion supports a persistent circling motion of the high-density spots around the trap. At high degrees of circle swimming, the swimmers become localized around the center of the trap, while hydrodynamic interactions seem to slightly counteract this effective confinement. The transition from the off-center towards the centered density distributions appears to be smooth, and we quantified it by introducing several corresponding order parameters. 

A long-term goal to extend the present theory would be the characterization of motility-induced phase separation into a dense clustered state and a surrounding low-density gas-like state \cite{theurkauff2012dynamic,fily2012athermal,bialke2013microscopic,buttinoni2013dynamical, palacci2013living,redner2013structure, cates2013active,stenhammar2013continuum,suma2014motility, speck2014effective,takatori2014swim, stenhammar2014phase,matas2014hydrodynamic, wysocki2014cooperative,fily2014freezing,wittkowski2014scalar, zottl2014hydrodynamics,cates2015motility, bialke2015active,speck2015dynamical,bialke2015negative, speck2016collective,richard2016nucleation,siebert2017phase}. This phenomenon was observed in particle-based simulations of active Brownian particles \cite{fily2012athermal,buttinoni2013dynamical,redner2013structure,suma2014motility,stenhammar2014phase, matas2014hydrodynamic,wysocki2014cooperative, fily2014freezing,zottl2014hydrodynamics, bialke2015negative,richard2016nucleation,siebert2017phase} and described by different statistical or continuum approaches \cite{bialke2013microscopic,cates2013active,stenhammar2013continuum,
speck2014effective,takatori2014swim,fily2014freezing, wittkowski2014scalar,speck2015dynamical}. So far, the effect of hydrodynamic interactions on this scenario has only rarely been addressed \cite{matas2014hydrodynamic,zottl2014hydrodynamics}. Our DDFT by construction contains self-propulsion driving the phase separation, steric interactions to avoid a collapse of the clustered state, and hydrodynamic interactions. \rev{In previous theoretical approaches, input for the density dependence of the swimming speed \cite{cates2013active} or for the front-back imbalance of the pair-correlation function \cite{bialke2013microscopic,speck2014effective,speck2015dynamical} was required to capture the phenomenon.} An interesting question for statistical theories and DDFT is whether such an input will further be necessary in the future, or whether the theories will provide it in a self-consistent way\rev{, as encouraged by a recent theoretical study \cite{hartel2017microscopic}.} Moreover, one could \rev{then analyze how the clustering behavior is influenced by the circular swimming paths.} 
Apart from that, in the future also the dynamic behavior of pure active swimming rotors \cite{nguyen2014emergent,van2016spatiotemporal,wykes2016dynamic} could be considered in an analogous statistical approach, including the induced hydrodynamic interactions between the rotors. 
\rev{We note that, in a different context, for reorienting the swimming motion, e.g., by external fields the consequences have been analyzed for the translational behavior and for the swim stress and pressure \cite{takatori2014swimstress, yan2015swim}.
Possibly, the latter quantities could also be extracted using our approach and explicit swimmer model.}
Another extension concerns the treatment of crystallization effects \cite{archer2014solidification} for active microswimmers including hydrodynamic interactions.

\ack %iopart

The authors thank Urs Zimmermann for helpful discussions on the simulation scheme and Gerhard N\"agele for helpful comments.
Support of this work by the Deutsche Forschungsgemeinschaft through the priority program SPP 1726 is acknowledged.

\appendix
\section{}
\label{sec:Appendix}

In our model, each swimmer consists of two force centers in the fluid in the vicinity of the swimmer body as shown in figure~\ref{fig:model}. 
To constitute a realistic microswimmer, no net force and no net torque may be exerted on the fluid. 

Since the two anti-parallel forces have the same magnitude $f$, the net force vanishes by construction. 
The individual torques $\vect{T}_\pm=\vect{r}^\pm \times (\pm f \uvec{n})$ caused by the two force centers of the swimmer can be calculated from the distance vectors $\vect{r}^\pm$ defined in (\ref{defplus}) and (\ref{defminus}). 
Thus, they read
	\begin{eqnarray}
		\vect{T}_+ = &f \left( \alpha L \uvec{n} \times \uvec{n} +  \gamma L \uvec{u} \times \uvec{n} \right) &= f\gamma L \uvec{u} \times \uvec{n}, \\
		\vect{T}_- = &-f \left( - (1-\alpha) L \uvec{n} \times \uvec{n} +  \gamma L \uvec{u} \times \uvec{n} \right)&= - f\gamma L \uvec{u} \times \uvec{n},
	\end{eqnarray}
	and cancel so that the net torque vanishes, as required.

\section*{References}

% \References

\bibliographystyle{iopart-num}
\bibliography{ddft_circle}

\end{document}